\documentclass[
 reprint,
 twocolumn,
 superscriptaddress,
amsmath,amssymb,
aps,
prb,
longbibliography
]{revtex4-1}

\usepackage[utf8]{inputenc}
\usepackage{graphicx}% Include figure files
\usepackage{dcolumn}% Align table columns on decimal point
\usepackage{bm}% bold math
\usepackage{hyperref}% add hypertext capabilities
\usepackage{dirtytalk}
\usepackage[dvipsnames]{xcolor}
\usepackage{multirow}
\usepackage{tabularx}
\usepackage{tabularray}
\usepackage{booktabs}

\makeatletter
\renewcommand{\fnum@figure}{{\bf Fig.\:\thefigure}}
\renewcommand{\fnum@table}{{\bf Table \thetable}}

\renewcommand{\@caption@fignum@sep}{$\:|$\nobreak\hskip.5em plus.2em\ignorespaces}
\renewcommand{\thefigure}{\arabic{figure}}
\renewcommand{\thetable}{\arabic{table}}
\makeatother

\begin{document}
\title{Quantum spin nematic phase in a square-lattice iridate} 

\author{Hoon~Kim}
\altaffiliation{These authors contributed equally to this work.}
\author{Jin-Kwang~Kim}
\altaffiliation{These authors contributed equally to this work.}
\author{Jimin~Kim}
\author{Hyun-Woo~J.~Kim}
\author{Seunghyeok~Ha}
\author{Kwangrae~Kim}
\author{Wonjun~Lee}
\affiliation{Center for Artificial Low Dimensional Electronic Systems, Institute for Basic Science (IBS), 77 Cheongam-Ro, Pohang 37673, South Korea}
\affiliation{Department of Physics, Pohang University of Science and Technology, Pohang 37673, South Korea}
\author{Jonghwan~Kim}
\affiliation{Center for Van der Waals Quantum Solids, Institute for Basic Science (IBS), Pohang, 37673, Korea}
\affiliation{Department of Materials Science and Engineering, Pohang University of Science and Technology, Pohang 37673, Korea}
\author{Gil~Young~Cho}
\affiliation{Center for Artificial Low Dimensional Electronic Systems, Institute for Basic Science (IBS), 77 Cheongam-Ro, Pohang 37673, South Korea}
\affiliation{Department of Physics, Pohang University of Science and Technology, Pohang 37673, South Korea}
\author{Hyeokjun~Heo}
\author{Joonho~Jang}
\affiliation{Department of Physics and Astronomy, Seoul National University, Seoul 08826, South Korea}
\author{J.~Strempfer}
\author{G.~Fabbris}
\author{Y.~Choi}
\author{D.~Haskel}
\author{Jungho~Kim}
\author{J.~W.~Kim}
\affiliation{Advanced Photon Source, Argonne National Laboratory, Argonne, Illinois 60439, USA}
\author{B. J. Kim}%
\email{bjkim6@postech.ac.kr}
\affiliation{Center for Artificial Low Dimensional Electronic Systems, Institute for Basic Science (IBS), 77 Cheongam-Ro, Pohang 37673, South Korea}
\affiliation{Department of Physics, Pohang University of Science and Technology, Pohang 37673, South Korea}

\date{\today}
\maketitle

{\bf Spin nematic (SN) is a magnetic analog of classical liquid crystals, a fourth state of matter exhibiting characteristics of both liquid and solid~\cite{Blu69,And84}. Particularly intriguing is a valence-bond SN~\cite{Mil17,Lau05,Sha06,Sat13}, in which spins are quantum entangled to form a multi-polar order without breaking time-reversal symmetry, but its unambiguous experimental realization remains elusive. Here, we establish a SN phase in the square-lattice iridate Sr$\bm_2$IrO$\bm_4$, which approximately realizes a pseudospin one-half Heisenberg antiferromagnet (AF) in the strong spin-orbit coupling limit~\cite{Kim08,Kim09,Jac09,Ber19}. Upon cooling, the transition into the SN phase at $\bm T_{\rm \bf C}$\,$\bm\approx$\,263 K is marked by a divergence in the static spin quadrupole susceptibility extracted from our Raman spectra, and concomitant emergence of a collective mode associated with the spontaneous breaking of rotational symmetries. The quadrupolar order persists in the antiferromagnetic (AF) phase below $\bm T_{\rm\bf N}$\,$\bm\approx$\,230 K, and becomes directly observable through its interference with the AF order in resonant x-ray diffraction, which allows us to uniquely determine its spatial structure. Further, we find using resonant inelastic x-ray scattering a complete breakdown of coherent magnon excitations at short-wavelength scales, suggesting a resonating-valence-bond-like quantum entanglement in the AF state~\cite{Pia15,Sha17}. Taken together, our results reveal a quantum order underlying the N\'eel AF that is widely believed to be intimately connected to the mechanism of high temperature superconductivity (HTSC)~\cite{And87,Pat06,Kei15}.
}

\begin{figure*}[hbt!]
\centering
\includegraphics[width=\textwidth]{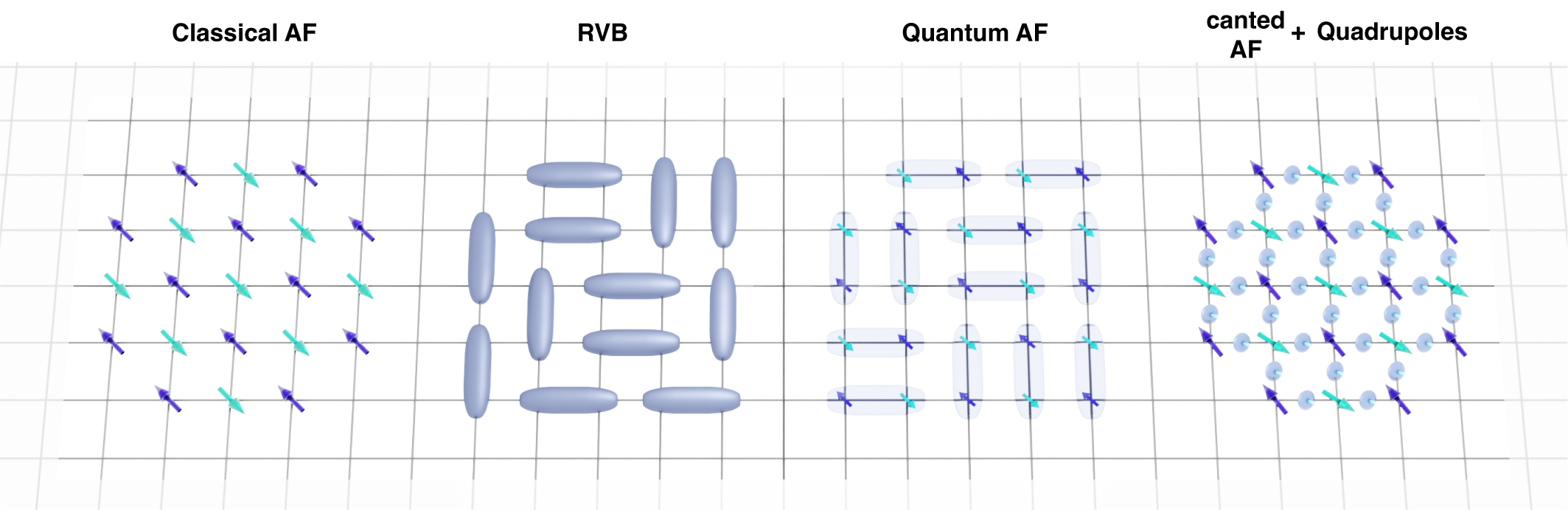}
\caption{\label{fig:1}{\bf Spin one-half moments on a square lattice.} In a strict 2D, AF order is not allowed at any finite temperature and spins are disordered but quantum entangled in a non-trivial way. For example, RVB is a state in which spins pair up to form singlets and fluctuate among numerous different configurations covering the square lattice with such dimers. Although most quasi-2D materials develop AF orders due to weak inter-layer couplings, spins still undergo strong quantum fluctuations. In a canted AF, quantum entanglement can manifest as an ordered spin quadrupole moment.  
}
\end{figure*}

With its relevance to HTSC in cuprates, spin one-half ($S$\,=\,1/2) Heisenberg model on a square lattice has been a subject of extensive research over the last several decades~\cite{Kei15,Pat06,Sca12,Ber19}. Although Mermin-Wagner theorem states that continuous symmetries cannot be spontaneously broken in two dimensions (2D), it is well established that N\'eel-type AF orders develop at low temperatures due to weak inter-layer couplings that form a 3D network in quasi-2D materials~\cite{Vak87}. Nevertheless, spins retain much of their properties in the disordered phase and undergo large quantum zero-point motions~\cite{Sin89,Col01}. The resulting ground state wave function is believed to embody a highly non-trivial structure akin to those in quantum spin liquids~\cite{Sav16,Zho17}. For instance, the resonant valence bond (RVB), a superposition of states in which spins pair up to form singlet ``valence bonds'', may have substantial overlap with the ground state~\cite{And87} (Fig.~\ref{fig:1}). Upon carrier doping, these singlets evolve into Cooper pairs in a prominent theory of HTSC~\cite{Pat06,And87}.  

What is not well known, however, is the fact that such quantum entanglement between a pair of nearest-neighbor (NN) spins can also manifest as an ordered spin quadrupole moment when spins are canted~\cite{Mil17,Lau05,Sha06,Sat13} (Fig.~\ref{fig:1}). Because quadrupole moment can be non-zero only for $S$\,$\geq$\,1, existence of a quadrupolar order in a $S$\,=\,1/2 system necessarily implies that spin pairs are entangled and have $S$\,=\,1 triplet components. Unlike singlets, however, a quadrupolar order is, in principle, measurable~\cite{Bar93,Sme13,Sav15} as it leads to spontaneous breaking of rotation symmetries~\cite{Koh19,Orl17}. In this Article, we show that a quadrupolar order coexists with the N\'eel AF order in Sr$_2$IrO$_4$, a single-layer ($n$=1) member of the Ruddlesden-Popper series iridates (Fig.~\ref{fig:2}a), which has received much attention due to its similarities with superconducting cuprates in their phenomenology~\cite{Kim08,Kim09,Jac09,Ber19}. Furthermore, we show that it persists above $T_{\rm N}$ and realizes a SN phase.

\section{Circular Dichroic \\ Resonant x-ray Diffraction}

\begin{figure*}
\centering
\includegraphics[width=0.75\textwidth]{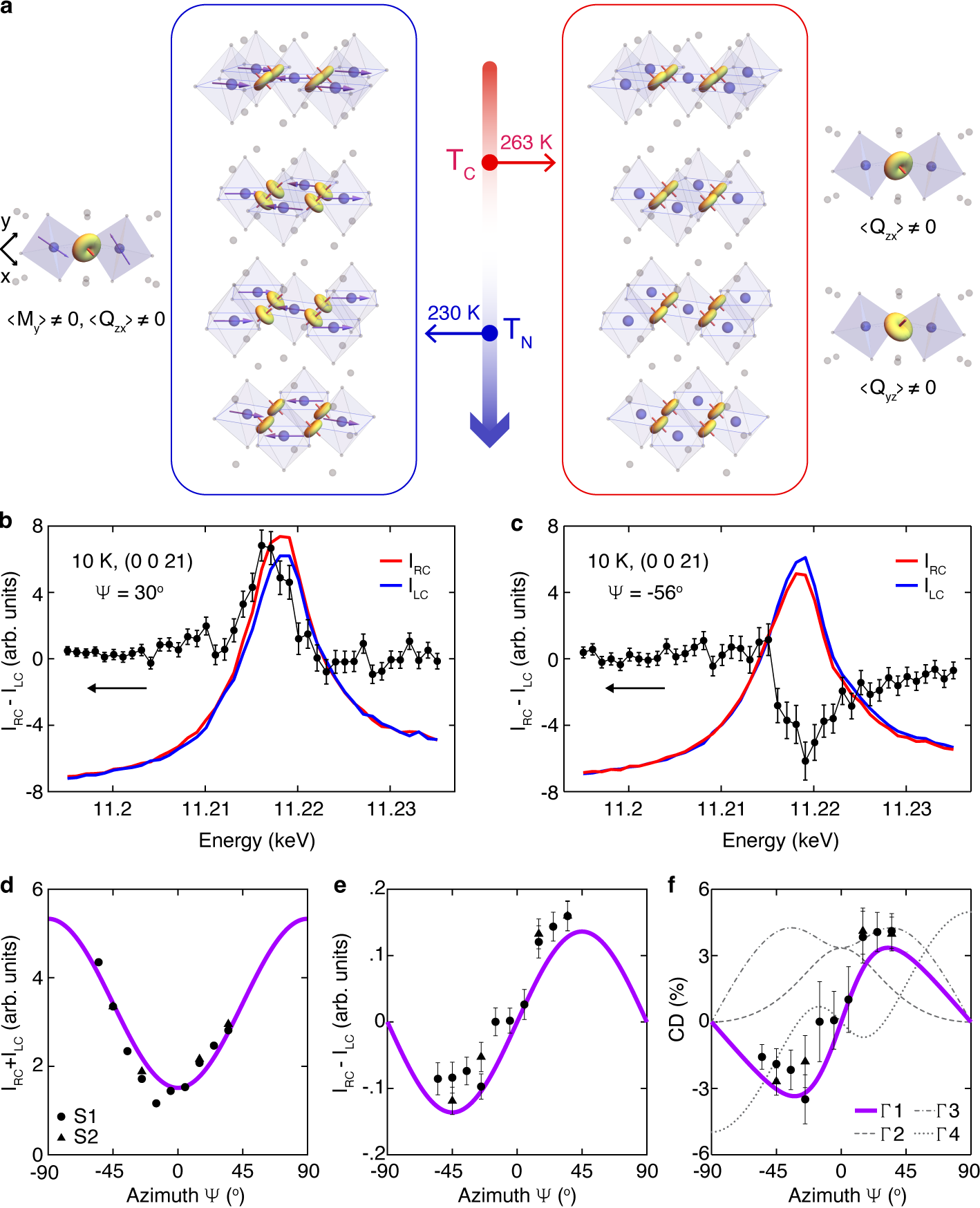}
\caption{\label{fig:2}{\bf Dipole-quadrupole interference in CD-RXD}. {\bf a,} A schematic for the N\'eel and SN orders overlaid on the crystal structure of Sr$_2$IrO$_4$ having the tetragonal $I4_1/acd$ space group. The magnetic and quadrupole moments mutually constrain their orientations below $T_N$ (left), whereas ($Q_{zx}$, $Q_{yz}$) transforms as an $E_{\rm g}$ doublet above $T_N$ (right). {\bf b,\,c,} Resonance profile of the circular dichroic signal at (0\,0\,21) magnetic reflection, measured at $\Psi$\,=\,30$^\circ$ (\textbf{b}) and $\Psi$\,=\,-56$^\circ$ (\textbf{c}). $I_{\rm RC}$ (red) and $I_{\rm LC}$ (blue) denote the diffraction intensity for right- and left-circularly polarized incident x-rays, respectively. {\bf d,\,e,} $I_{\rm SUM}\,\equiv\,I_{\rm RC}+I_{\rm LC}$ (\textbf{d}) and $I_{\rm DIFF}\,\equiv\,I_{\rm RC}-I_{\rm LC}$ (\textbf{e}) at the resonance for the full range of $\Psi$ measured. S1 and S2 are obtained from two independent measurements. \textbf{f,} Comparison of $I_{\rm CD}\,\equiv\,I_{\rm DIFF}/I_{\rm SUM}$ to the simulations for different irreducible representations. The best fit to the data using $\Gamma_1$ profile is shown in \textbf{d-f}, and its structure is shown in \textbf{a}. }
\end{figure*}

We use resonant x-ray diffraction (RXD) to show the coexistence of a quadruoplar and a dipolar (canted AF) order (Fig.~\ref{fig:2}a). 
The dipole-quadrupole interference contribution to the RXD intensity can be isolated by measuring circular dichroic (CD) signal of a magnetic Bragg reflection defined as 

\begin{equation}
I_{\textrm{CD}} = \frac{I_\textrm{DIFF}}{I_\textrm{SUM}}=\frac{I_{\textrm{LL}} + I_{\textrm{LR}} - I_{\textrm{RR}} - I_{\textrm{RL}}}{I_{\textrm{LL}} + I_{\textrm{LR}} + I_{\textrm{RR}} + I_{\textrm{RL}}}\;,
\end{equation}
\\
\noindent where $I_{\textrm{L(R)L(R)}}$ denotes the intensity of left (right) circular polarized incident x-ray scattered into left (right) circular outgoing x-ray. This definition of $I_{\textrm{CD}}$ reflects the fact that the scattered x-ray polarization is not resolved in our experiment.
It is straightforward to show that the difference spectrum ($I_{\textrm{DIFF}}$) in our particular case can be expressed as (see Supplementary Note S1)

\begin{equation}
\begin{split}
I_{\textrm{DIFF}}\propto F_1^Me_1^A F_2^Qe_2^S+F_1^Qe_1^S F_2^Me_2^A \\
- F_2^Me_1^A F_1^Qe_2^S - F_2^Qe_1^S F_1^Me_2^A\;,
\end{split}
\end{equation}
\\
\noindent
where $F_{1,2}^Q$ and $F_{1,2}^M$ couple to symmetric ($e_1^S$) and antisymmetric ($e_2^A$) component of the polarization tensor $e_{\alpha\beta}\equiv\epsilon^{\prime*}_\alpha\epsilon_\beta$, respectively, and represent the complex structure factor ($F$\,$\equiv$\,$F_1$+$i F_2$) associated with time-reversal-even quadrupoles and time-reversal-odd dipoles, respectively. Thus, CD arises from the interference between dipolar and quadrupolar scatterings in the electric-dipole-electric-dipole ($E1$-$E1$) process of RXD. In Supplementary Note S2, we argue that this is the unique explanation for the CD in our case among all known mechanisms. 

Figures~\ref{fig:2}b and \ref{fig:2}c show the representative resonance profile of the (0\,0\,21) reflection, arising from the net FM moment due to canting of the spins~\cite{Kim09}, at two different azimuth angles ($\Psi$), defined as the angle between the crystallographic $a$-axis and the vertical scattering plane. $I_{\textrm{DIFF}}$ is non-zero at and close to the resonance, and positive (negative) at $\Psi$\,=\,30$^\circ$ ($\Psi$\,=\,-56$^\circ$). Figures~\ref{fig:2}d and~\ref{fig:2}e show $I_{\textrm{SUM}}$ and $I_{\textrm{DIFF}}$ at the resonance for the full range of $\Psi$ measured. The $\Psi$-dependence of $I_{\textrm{CD}}$ (Fig.~\ref{fig:2}f) allows us to determine the spatial structure and symmetry of the quadrupolar order. To systematically find all symmetry-allowed $\mathbf{q}$\,=\,0 bond-centered quadrupole structures (see Supplementary Note S3), we first note that the reflection appearing at (0\,0\,21) implies that both the dipolar and the quadrupolar orders break the body-center translation symmetry. Such structures are represented by one of the four irreducible representations (IRs) $\Gamma_1$, $\Gamma_2$, $\Gamma_3$, and $\Gamma_4$, which are all 2D. For example, it is known that the magnetic structure shown in Fig.~\ref{fig:2}a belongs to $\Gamma_1$ IR~\cite{Ye13}.

While there are many different possible quadrupole structures (see Supplementary Note S3), there are only a few that have non-zero structure factors, which involve $yz$/$zx$ quadrupoles for $\Gamma_1$ and $\Gamma_2$, $xy$ for $\Gamma_3$, and $x^2-y^2$\,/\,$3z^2-r^2$ for $\Gamma_4$. In Fig.~\ref{fig:2}f, the $\Psi$-dependence of $I_{\textrm{CD}}$ is simulated for each of these structures and compared with the data. We find that only the $\Gamma_1$ structure, depicted in Fig.~\ref{fig:2}a, is consistent with the experimental data, whose key feature is the sign change at $\Psi$\,=\,0$^\circ$. The total scattering amplitude is given by

\begin{align}
&F = F^M + r F^Q \\
  &=  i\left(
    \begin{matrix}
    0 & 0 & -c M_y\\
    0 & 0 & c^* M_x\\
   c M_y & -c^* M_x & 0
    \end{matrix}
    \right)
   +r
     \left(\begin{matrix}
    0 & 0 & c Q_{zx}\\
    0 & 0 & c^* Q_{yz}\\
    c Q_{zx} & c^* Q_{yz}& 0
    \end{matrix}
    \right), \notag
\end{align}
\\
\noindent
where $\mathbf{M}$\,=\,($M_x$, $M_y$) and $\mathbf{Q}$\,=\,($Q_{yz}$, $Q_{zx}$)  are 2D real basis vectors (whose transformation matrices are listed in Supplementary Note S3) for $\Gamma_1$  representing the net FM component of the magnetic order and the quadrupolar order, respectively, normalized to one when they have the maximal values; and $c$\,=\,$4(1-i)$ is a constant arising from the structure factor. $r$ is a dimensionless factor accounting for the fact that quadrupoles only become visible by high-order RXD processes (see Supplementary Note S4 for possible microscopic mechanisms).  

For the magnetic order, it is known that the easy axis is along $<$100$>$~\cite{Por19}, and the fact that $I_{\textrm{SUM}}$ has a minimum at $\Psi$\,=\,0$^\circ$ means that the magnetic domain being measured has the FM (AF) component along the $b$-($a$-)axis, and hence $M_x$\,=\,0. As we shall see below, fixing the direction of $\mathbf{M}$ along $y$ constrains $Q_{yz}$ to be zero. Thus, it follows that

\begin{equation}
    I_{\textrm{CD}} = \frac{ (r Q_{zx}/M_{y}) \;\sin{\alpha} \sin{2\Psi}}{1-\cos{2\alpha}\cos{\Psi}^2+(r Q_{zx}/M_y )^2 \sin{\Psi}^2},
\end{equation}
\\
\noindent
where $\alpha$ is incidence angle to the sample surface ($\sim27^\circ$). Thus, the data can be fit with $r {Q_{zx}}/M_y$ as the only fitting parameter, and the best fit is obtained when $r {Q_{zx}}/M_y$\,$\approx$\,0.04 (Fig.~\ref{fig:2}f). With a quantitative estimate of the $r$ factor, which is beyond the scope of this work, one can in principle compare the relative magnitudes of $Q_{zx}$ and $M_y$. 

\section{Two-site \textbf{\textit{S}}\,=\,1/2 model}

\begin{table}[b!]
\caption{\textbf{Quadrupole and dipole moments from the two-site ${\bm S}$=1/2 model calculation.}
}
\centering
\begin{ruledtabular}
\begin{tabular}{c c c}
\addlinespace[4pt]
\textbf{Type} & \textbf{Order} & \;\;\; \textbf{Expression} \\
\addlinespace[6pt]\hline\addlinespace[4pt]
 AF & $\langle N_x \rangle$ & \;\;\; ${\sin 2\theta}{\cos \phi_c}$ \\ \addlinespace[4pt]
 FM & $\langle M_y \rangle$ & \;\;\; ${\sin 2\theta} {\sin \phi_c}$ \\ \addlinespace[4pt]
\hline \addlinespace[4pt]
\footnotesize{Quadrupole} & $\langle Q_{zx} \rangle$ & $\frac{1}{2}\left({\cos 2\theta} \,{\sin 2\phi} \, {\sin \phi_c}\right)$ \\ \addlinespace[4pt]
Singlet & $\langle\text{-}\mathbf{S}^1 \cdot \mathbf{S}^2 \text{-} \frac{1}{4}\rangle$ & $\frac{1}{2}\left({\cos 2\theta}\,{\cos 2\phi}\,{\cos^2\phi_c} - \sin^2 \phi_c\right)$ \\ \addlinespace[4pt]
\end{tabular}
\end{ruledtabular}
\label{tab:1}
\end{table}

When a quadrupolar order coexists with a dipolar order, its allowed structure is constrained by the orientation of the dipolar order. For example, for a single $S$\,=\,1 spin, it is easy to show that if its dipole moment is along $z$, then $yz$ and $zx$ quadrupole moments are zero. For the present case, we proceed with a simple two-site $S$\,=\,1/2 model (see Supplementary Note S5). An arbitrary wavefunction for a pair of NN spins can be expressed in the singlet-triplet basis ($s$,\,$T_x$,\,$T_y$,\,$T_z$) in terms of two parameters $\theta$ and $\phi$ as $\mathbf{u}$\,+\,$i\mathbf{v}$, where

\begin{equation}
\begin{split}
    \mathbf{u}&\,=\,\cos\theta (\cos{\phi}\cos{\phi_c}, -\sin{\phi},\, 0,\, \cos{\phi}\sin{\phi_c}) \;,\\
    \mathbf{v}&\,=\,\sin\theta\,(\sin{\phi}\cos{\phi_c},\quad \cos{\phi},\, 0,\, \sin{\phi}\sin{\phi_c})\;,
\end{split}
\end{equation}
\\
\noindent
under the constraints that the AF(FM) component of the ordered moment is along $a$($b$) axis, and the canting angle $\phi_c$\,$\approx$\,12$^\circ$; i.e., $\langle N_y \rangle$\,=\,$\langle N_z \rangle$\,=\,$\langle M_x \rangle$\,=\,$\langle M_z \rangle$\,=\,0 and $\langle M_y \rangle/\langle N_x \rangle$\,=\,$\tan \phi_c$, (where $N_\alpha$\,$\equiv$\,$S^1_\alpha-S^2_\alpha$ and $M_\alpha$\,$\equiv$\,$S^1_\alpha+S^2_\alpha$).

\begin{figure*}
\centering
\includegraphics[width=0.75\textwidth]{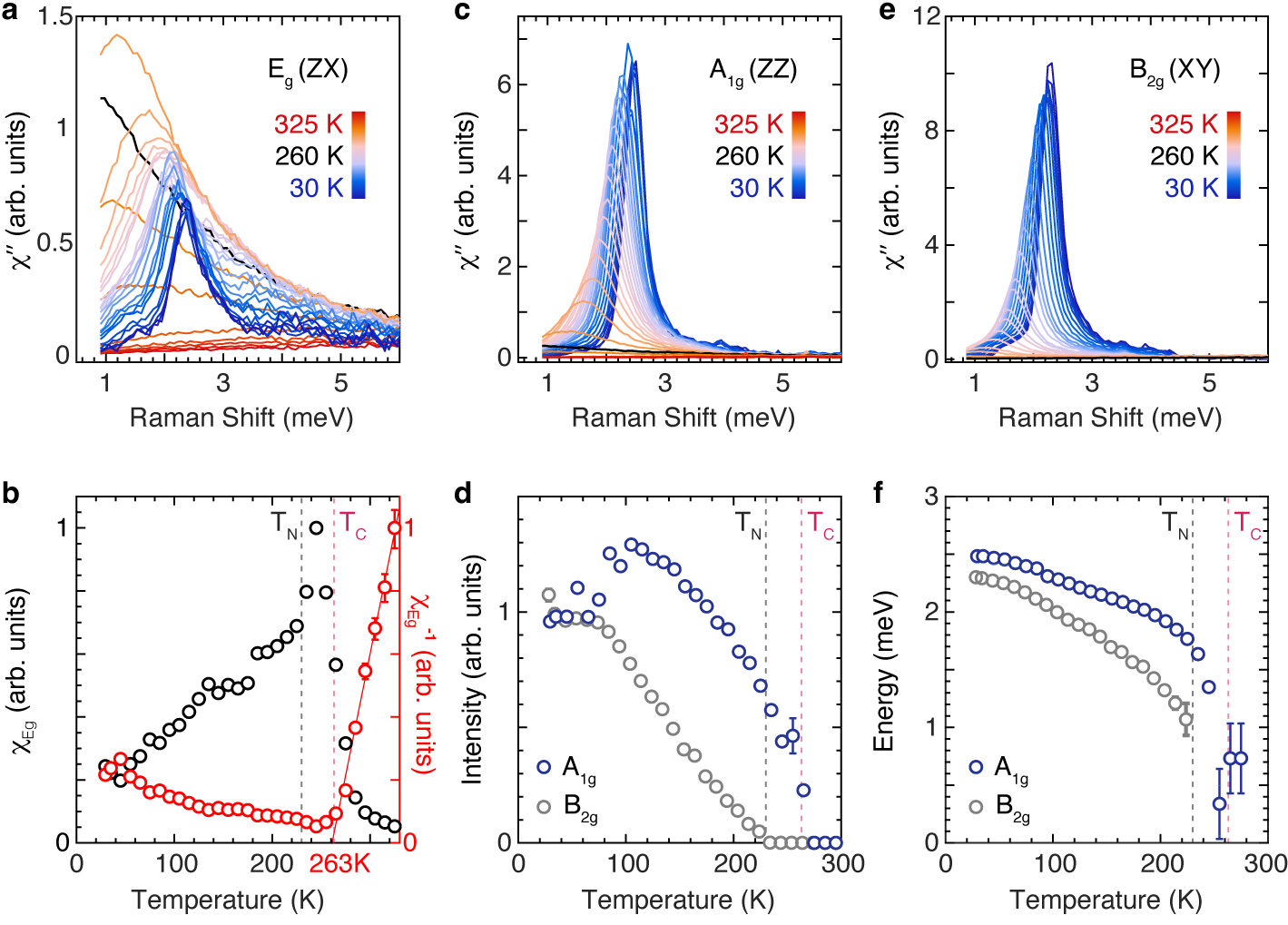}
\caption{\label{fig:3}{\bf Phase transition into the spin-nematic phase}. {\bf a,} Raman spectra in the $E_{\rm g}$ symmetry channel measured on cooling. {\bf b,} The real part of the uniform static susceptibility (black) obtained by integrating the $E_{\rm g}$ Raman conductivity $\chi''/\omega$. The inverse susceptibility (red) exhibits a Curie-Weiss behaviour with $T_C\,\simeq$\,263\,K. {\bf c,\,e,} $A_{\rm 1g}$ phase mode emerges at $T_C$ (\textbf{c}), and the $B_{\rm 2g}$ magnon below $T_{\rm N}$ (\textbf{e}). {\bf d,\,f,} Temperature dependence of the intensities (\textbf{d}) and energies (\textbf{f}) of $A_{\rm 1g}$ and $B_{\rm 2g}$ modes extracted by fitting the spectra to Fano profiles~$I(\omega)\,=\,\frac{I_0}{\Gamma(1-q^2)}(1-\frac{(q+\epsilon)^2}{1+\epsilon^2})$, where $\epsilon\,=\,(\omega-\omega_0)/\Gamma$, $\Gamma$ is the linewidth, and $q$ is the asymmetry parameter. }
\end{figure*}

Table~1 compares the magnitudes of magnetic and nonmagnetic orders calculated as functions of $\theta$ and $\phi$. First, we find that $\langle Q_{xy} \rangle$\,=\,$\langle Q_{yz} \rangle$\,=\,0 for all values of $\theta$ and $\phi$, but $\langle Q_{zx}\rangle$ can be nonzero only when $\phi_c$ is nonzero. This suggests that Dzyloshinskii-Moriya (DM) interaction plays an important role in stabilizing the quadrupolar order, since the spin canting arises from the same origin~\cite{Jac09}. Second, it is clear that $\theta$ parameterizes the competition between the magnetic and nonmagnetic sectors. For example, at $\theta$\,=\,$\frac{\pi}{4}$ the magnetic moment fully saturates and $\langle Q_{zx} \rangle$\,=\,0. Third, in the non-magnetic sector $\phi$ controls the magnitude of $\langle Q_{zx} \rangle$, which anti-correlates with that of the dimer correlation  $\langle -\mathbf{S}^1 \cdot \mathbf{S}^2 - \frac{1}{4} \rangle$, which takes the values $\frac{1}{2}$, 0, $-\frac{1}{2}$ for the singlet, N\'eel AF state, and triplets, respectively.  
In other words, just as the net FM moment has its origin in the canting of AF spins, $\langle Q_{zx} \rangle$ arises from the underlying singlet correlation by ``canting'' in angle $\phi$. This raises an intriguing possibility that a SN may arise from a smooth deformation of a RVB state that has non-zero singlet correlation for every NN pair spins. For instance, recent studies find that rotating one of the spins by 180$^\circ$ in every dimer in a singlet RVB leads to a triplet RVB, which inherits some of the generic properties of the RVBs, such as quasiparticles with fractional  quantum numbers~\cite{Kon22,Shi09}. 

\section{Raman Spectroscopy}
Next, we use Raman spectroscopy, which has been suggested as a sensitive probe for SN~\cite{Mil17}, to show that the quadrupolar order persists above $T_{\rm N}$, armed with the information that the order should appear in the YZ or ZX polarization channel, which correspond to $E_g$ symmetry channel of the tetragonal $D_{4h}$ point group. This requires the laser beam to be incident on the side surface of a thin plate-like crystal, and thus has not been measured in previous Raman studies on Sr$_2$IrO$_4$~\cite{Cet12,Gim16,Gre16}. Figure~\ref{fig:3}a shows the Raman spectra measured in ZX scattering geometry. Upon cooling down from $T$\,$=$\,325 K, we observe a broad quasi-elastic feature developing with its tail extending up to $\lesssim$\,10\,meV, which is indicative of slow fluctuations of spins (Fig.~\ref{fig:3}a). Integrating the Raman conductivity $\chi''/\omega$ over a sufficiently large energy window (0.85\,meV$\sim$ 23.4\,meV), we obtain in Fig.~\ref{fig:3}b the real part of the static susceptibility, which follows the Curie-Weiss temperature dependence with $T_{\rm C}$\,$\approx$\,263 K. 

Concomitantly, an $A_{1g}$ mode emerges (Fig.~\ref{fig:3}c and Extended Data Fig.~\ref{fig:ed1}), which is barely visible above our instrumental low-energy cutoff $\approx$\,0.85 meV at $T_{\rm C}$, but upon further cooling becomes well resolved as the peak moves to higher energies. We interpret this peak as the phase mode associated with the quadrupolar order, analogous to spin-wave modes in the AF phase. The order parameter has $E_g$ symmetry, which can be continuously rotated about the $c$-axis. This allows the phase mode to have the energy (1$\sim$2 meV) an order of magnitude smaller than the temperature scale of $T_{\rm C}$ ($\sim$20 meV), in accordance with the Goldstone theorem. Together with the divergence in the static susceptibility, this collective excitation constitutes direct evidence of a thermodynamic phase transition at $T_C$\,$\sim$\,263 K, well above the AF transition temperature $T_{\rm N}$\,$\sim$\,230 K.

The $A_{1g}$ mode persists as the sample is cooled down through $T_{\rm N}$ (Figs.~\ref{fig:3}c and \ref{fig:3}d), at which temperature the $B_{2g}$ single-magnon mode emerges (Fig.~\ref{fig:3}e). The $A_{1g}$ mode becomes gapped close to $T_{\rm N}$ as the AF order further reinforces the breaking of the rotational symmetry (Fig.~\ref{fig:3}f). The intensity of the $A_{1g}$ mode continues to grow in the AF phase (Fig.~\ref{fig:3}d), implying coexistence of dipolar and quadrupolar orders. As the AF order sets in, however, the stacking pattern of the quadrupoles changes, which is governed by weak inter-layer interactions of $\lesssim$10 $\mu$eV energy scales, four orders of magnitudes smaller than those for intra-layer interactions~\cite{Por19} (Fig.~\ref{fig:2}a). We note that the $A_{1g}$ mode is almost insensitive to the onset of the dipolar order, indicating that spins entanglement remains intact below $T_{\rm N}$. 

\section{Polarization-resolved \\ Resonant inelastic x-ray scattering}
\begin{figure*}[ht!]
\centering
\includegraphics[width=0.75\textwidth]{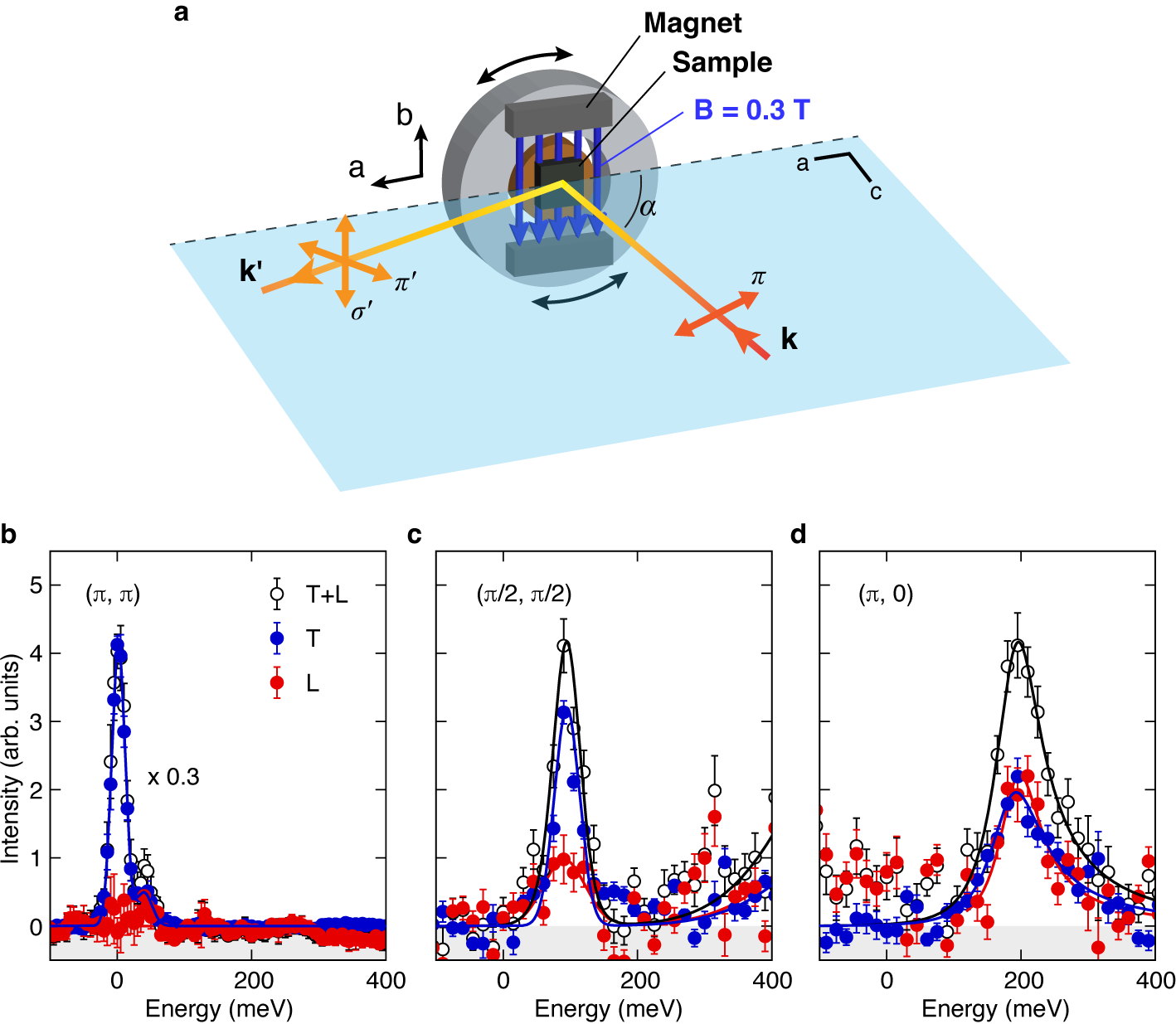}
\caption{\label{fig:4}{\bf Complete breakdown of magnon at short wavelength scales}. \textbf{a,} A schematic of the RIXS geometry. X-rays with $\pi$ polarization are incident to the sample with the angle $\alpha \approx 75^{\circ}$, and scattered x-rays are collected without polarization analysis. A single crystal of Sr$_2$IrO$_4$ is placed between two permanent magnets that apply a magnetic field of 0.3~T along the $b$-axis ($a$-axis) at the sample position aligning the magnetic domains, i.e., $\mathbf{M}$\,$\parallel$\,$\hat{b}$ ($\mathbf{M}$\,$\parallel$\,$\hat{a}$). The direction of the applied field can be changed by rotating the outer disk on which the magnets are mounted while the sample stays fixed. The angle between the incident and outgoing x-rays is fixed close to 90$^{\circ}$ to suppress elastic Thomson scattering. {\bf b-d,} RIXS spectra for $(\pi, \pi)$ (\textbf{b}), $(\pi/2, \pi/2)$ (\textbf{c}), and $(\pi, 0)$ (\textbf{d}). The transverse mode (T) and the longitudinal mode (L) are displayed along with the sum of the two (T+L). Solid lines are guides to the eye. Each components are extracted from the raw spectra measured in different magnetic field directions with a fixed scattering geometry (see Extended Data Table~\ref{tab:ed1} and Supplementary Note S6).}
\end{figure*}

If the quadrupolar order comprises a substantial part of the ground state (small $\theta$ limit in Eq.~5), excitation spectra must exhibit a qualitative departure from the classical spin-wave theory (SWT), which largely reproduces the experimental spectra of cuprate square-lattice AFs \cite{Aep89,Col01,Bra14,Pen17}. Figure~\ref{fig:4} shows the RIXS spectra for the spin components transverse (T) and longitudinal (L) to the ordered moment direction, which we resolve by measuring the spectra for two different magnetic moment directions aligned using a small permanent magnet (see Extended Data Table~\ref{tab:ed1} and Supplementary Note S6). At the ordering wavevector ($\pi$, $\pi$) (Fig.~\ref{fig:4}b), the spectral weight is mostly in the T channel, consistent with the SWT description, which expects a divergent intensity for a gapless Goldstone mode. The T magnon mode is also clearly visible at ($\pi/2$, $\pi/2$), although a significant weight is transferred to the L channel (Fig.~\ref{fig:4}c). By contrast, at ($\pi$, 0) the spectrum is completely isotropic and shows no sharp feature that can be identified as a magnon mode (Fig.~\ref{fig:4}d). 

In the cuprates, the `($\pi$,0) anomaly' is ubiquitously observed across many different materials~\cite{Chr07,Hea10,Pia15,Mar22,Tsy09,Tsy10}, and from its isotropic nature the continuum has been interpreted as deconfined fractional quasiparticles (spinons)~\cite{Pia15,Sha17}. We note, however, that the T mode still constitutes the dominant part of the total intensity in most cases. To the best of our knowledge, the complete loss of the coherent magnon intensity is unprecedented, and renders alternative multi-magnon scenarios~\cite{Pow18} unlikely in our case. Rather, attributing the isotropic continuum to deconfined spinons in turn suggests that the quadrupolar order may be a byproduct of a RVB-like state.

\section{Discussion}

Theoretically, SNs are predicted to arise in certain spin models with competing interactions~\cite{Blu69,Che71,Cha91,Lau06,Zhi10}. For example, in a square-lattice, four-spin exchange ($J_c$) competing with ferromagnetic (FM) NN Heisenberg exchange ($J$) can lead to a SN~\cite{Lau05,Sha06}. In this regard, 5$d$ transition-metal oxides, classified as `weak' Mott insulators stabilized by strong spin-orbit coupling, may be a promising ground to search for a SN, because the large spatial extent of the 5$d$ orbitals can bring about strong competing interactions with NN Heisenberg interactions~\cite{Wit14,Ber19}. Indeed, presence of interaction terms beyond NN in Sr$_2$IrO$_4$ is clear from its steep spin-wave dispersion along the magnetic zone-boundary~\cite{Kim12}, which can be equally well fitted with a model including $J_c$ or further-neighbor couplings ($J_2$ and $J_3$). Further, our study suggests that the large canting angle of the spins resulting in a sizable net FM moment in each layer may be favourable for stabilizing the SN phase even in an N\'eel AF.   

The fact that $L$ edges of 5$d$ transition-metal elements are in the hard x-ray regime with access to a wide region in the momentum space is also advantageous for direct detection of spin quadrupoles, as x-rays become sensitive to quadrupoles under resonance conditions~\cite{Sav15,Sme13,Mcm01}. Most other experimental probes are insensitive to spin quadrupoles, and evidence for a SN has so far been indirectly provided from thermodynamic~\cite{Mou12,Koh19} and nuclear magnetic resonance measurements~\cite{Orl17}. In Sr$_2$IrO$_4$, evidence for a symmetry breaking order above $T_{\rm N}$ has been found in studies using second harmonic generation~\cite{Sey20}, inelastic neutron scattering~\cite{Jeo17}, and magnetic torque measurements~\cite{Mur21}. These studies suggest loop currents as the possible order. While our result is inconsistent with such a time-reversal symmetry breaking order (see Extended Data Fig.~\ref{fig:ed2}), it is possible that these experiments are probing yet another order of different nature.

Our results vividly demonstrate that mulitple orders are intertwinned even in the Mott insulating phase. The discovery of the SN phase is not only significant on its own, but also pave a new pathway to investigate the entanglement structure of quantum spins and the possible connection between SN and RVB through resonant x-ray scattering processes~\cite{Sav15}. The quantum entanglement between NN spins manifests as a spin quadrupolar order, and possibly accounts for the complete loss of coherent magnons at momentum ($\pi$,\,0). This in turn suggests that the quadrupolar order may be exploited to follow the singlets formed by NN spins. Along this direction, if quadrupolar orders can be detected through charge-quadrupole interference, their temperature and doping evolution can be followed in the absence of a magnetic order, with the ultimate goal of elucidating the relationship between the role of magnetic correlations in myriad exotic phases displayed by doped Mott insulators.

%\bibliography{natbib}

\begin{thebibliography}{56}%
\makeatletter
\providecommand \@ifxundefined [1]{%
 \@ifx{#1\undefined}
}%
\providecommand \@ifnum [1]{%
 \ifnum #1\expandafter \@firstoftwo
 \else \expandafter \@secondoftwo
 \fi
}%
\providecommand \@ifx [1]{%
 \ifx #1\expandafter \@firstoftwo
 \else \expandafter \@secondoftwo
 \fi
}%
\providecommand \natexlab [1]{#1}%
\providecommand \enquote  [1]{``#1''}%
\providecommand \bibnamefont  [1]{#1}%
\providecommand \bibfnamefont [1]{#1}%
\providecommand \citenamefont [1]{#1}%
\providecommand \href@noop [0]{\@secondoftwo}%
\providecommand \href [0]{\begingroup \@sanitize@url \@href}%
\providecommand \@href[1]{\@@startlink{#1}\@@href}%
\providecommand \@@href[1]{\endgroup#1\@@endlink}%
\providecommand \@sanitize@url [0]{\catcode `\\12\catcode `\$12\catcode
  `\&12\catcode `\#12\catcode `\^12\catcode `\_12\catcode `\%12\relax}%
\providecommand \@@startlink[1]{}%
\providecommand \@@endlink[0]{}%
\providecommand \url  [0]{\begingroup\@sanitize@url \@url }%
\providecommand \@url [1]{\endgroup\@href {#1}{\urlprefix }}%
\providecommand \urlprefix  [0]{URL }%
\providecommand \Eprint [0]{\href }%
\providecommand \doibase [0]{http://dx.doi.org/}%
\providecommand \selectlanguage [0]{\@gobble}%
\providecommand \bibinfo  [0]{\@secondoftwo}%
\providecommand \bibfield  [0]{\@secondoftwo}%
\providecommand \translation [1]{[#1]}%
\providecommand \BibitemOpen [0]{}%
\providecommand \bibitemStop [0]{}%
\providecommand \bibitemNoStop [0]{.\EOS\space}%
\providecommand \EOS [0]{\spacefactor3000\relax}%
\providecommand \BibitemShut  [1]{\csname bibitem#1\endcsname}%
\let\auto@bib@innerbib\@empty
%</preamble>
\bibitem [{\citenamefont {Blume}\ and\ \citenamefont {Hsieh}(1969)}]{Blu69}%
  \BibitemOpen
  \bibfield  {author} {\bibinfo {author} {\bibfnamefont {M.}~\bibnamefont
  {Blume}}\ and\ \bibinfo {author} {\bibfnamefont {Y.~Y.}\ \bibnamefont
  {Hsieh}},\ }\bibfield  {title} {\enquote {\bibinfo {title} {{Biquadratic
  Exchange and Quadrupolar Ordering}},}\ }\href {\doibase 10.1063/1.1657616}
  {\bibfield  {journal} {\bibinfo  {journal} {J. Appl. Phys.}\ }\textbf
  {\bibinfo {volume} {40}},\ \bibinfo {pages} {1249–1249} (\bibinfo {year}
  {1969})}\BibitemShut {NoStop}%
\bibitem [{\citenamefont {Andreev}\ and\ \citenamefont
  {Grishchuk}(1984)}]{And84}%
  \BibitemOpen
  \bibfield  {author} {\bibinfo {author} {\bibfnamefont {AF}~\bibnamefont
  {Andreev}}\ and\ \bibinfo {author} {\bibfnamefont {IA}~\bibnamefont
  {Grishchuk}},\ }\bibfield  {title} {\enquote {\bibinfo {title} {{Spin
  Nematics}},}\ }\href {http://www.jetp.ras.ru/cgi-bin/dn/e_060_02_0267.pdf}
  {\bibfield  {journal} {\bibinfo  {journal} {Sov. Phys. JETP}\ }\textbf
  {\bibinfo {volume} {60}},\ \bibinfo {pages} {267} (\bibinfo {year}
  {1984})}\BibitemShut {NoStop}%
\bibitem [{\citenamefont {Mila}(2017)}]{Mil17}%
  \BibitemOpen
  \bibfield  {author} {\bibinfo {author} {\bibfnamefont {Frédéric}\
  \bibnamefont {Mila}},\ }\bibfield  {title} {\enquote {\bibinfo {title}
  {{Closing in on a magnetic analog of liquid crystals}},}\ }\href {\doibase
  10.1103/Physics.10.64} {\bibfield  {journal} {\bibinfo  {journal} {Physics}\
  }\textbf {\bibinfo {volume} {10}},\ \bibinfo {pages} {64} (\bibinfo {year}
  {2017})}\BibitemShut {NoStop}%
\bibitem [{\citenamefont {L\"auchli}\ \emph {et~al.}(2005)\citenamefont
  {L\"auchli}, \citenamefont {Domenge}, \citenamefont {Lhuillier},
  \citenamefont {Sindzingre},\ and\ \citenamefont {Troyer}}]{Lau05}%
  \BibitemOpen
  \bibfield  {author} {\bibinfo {author} {\bibfnamefont {A.}~\bibnamefont
  {L\"auchli}}, \bibinfo {author} {\bibfnamefont {J.~C.}\ \bibnamefont
  {Domenge}}, \bibinfo {author} {\bibfnamefont {C.}~\bibnamefont {Lhuillier}},
  \bibinfo {author} {\bibfnamefont {P.}~\bibnamefont {Sindzingre}}, \ and\
  \bibinfo {author} {\bibfnamefont {M.}~\bibnamefont {Troyer}},\ }\bibfield
  {title} {\enquote {\bibinfo {title} {{Two-Step Restoration of SU(2) Symmetry
  in a Frustrated Ring-Exchange Magnet}},}\ }\href {\doibase
  10.1103/PhysRevLett.95.137206} {\bibfield  {journal} {\bibinfo  {journal}
  {Phys. Rev. Lett.}\ }\textbf {\bibinfo {volume} {95}},\ \bibinfo {pages}
  {137206} (\bibinfo {year} {2005})}\BibitemShut {NoStop}%
\bibitem [{\citenamefont {Shannon}\ \emph {et~al.}(2006)\citenamefont
  {Shannon}, \citenamefont {Momoi},\ and\ \citenamefont {Sindzingre}}]{Sha06}%
  \BibitemOpen
  \bibfield  {author} {\bibinfo {author} {\bibfnamefont {Nic}\ \bibnamefont
  {Shannon}}, \bibinfo {author} {\bibfnamefont {Tsutomu}\ \bibnamefont
  {Momoi}}, \ and\ \bibinfo {author} {\bibfnamefont {Philippe}\ \bibnamefont
  {Sindzingre}},\ }\bibfield  {title} {\enquote {\bibinfo {title} {{Nematic
  Order in Square Lattice Frustrated Ferromagnets}},}\ }\href {\doibase
  10.1103/PhysRevLett.96.027213} {\bibfield  {journal} {\bibinfo  {journal}
  {Phys. Rev. Lett.}\ }\textbf {\bibinfo {volume} {96}},\ \bibinfo {pages}
  {027213} (\bibinfo {year} {2006})}\BibitemShut {NoStop}%
\bibitem [{\citenamefont {Sato}\ \emph {et~al.}(2013)\citenamefont {Sato},
  \citenamefont {Hikihara},\ and\ \citenamefont {Momoi}}]{Sat13}%
  \BibitemOpen
  \bibfield  {author} {\bibinfo {author} {\bibfnamefont {Masahiro}\
  \bibnamefont {Sato}}, \bibinfo {author} {\bibfnamefont {Toshiya}\
  \bibnamefont {Hikihara}}, \ and\ \bibinfo {author} {\bibfnamefont {Tsutomu}\
  \bibnamefont {Momoi}},\ }\bibfield  {title} {\enquote {\bibinfo {title}
  {{Spin-Nematic and Spin-Density-Wave Orders in Spatially Anisotropic
  Frustrated Magnets in a Magnetic Field}},}\ }\href {\doibase
  10.1103/PhysRevLett.110.077206} {\bibfield  {journal} {\bibinfo  {journal}
  {Phys. Rev. Lett.}\ }\textbf {\bibinfo {volume} {110}},\ \bibinfo {pages}
  {077206} (\bibinfo {year} {2013})}\BibitemShut {NoStop}%
\bibitem [{\citenamefont {Kim}\ \emph {et~al.}(2008)\citenamefont {Kim},
  \citenamefont {Jin}, \citenamefont {Moon}, \citenamefont {Kim}, \citenamefont
  {Park}, \citenamefont {Leem}, \citenamefont {Yu}, \citenamefont {Noh},
  \citenamefont {Kim}, \citenamefont {Oh}, \citenamefont {Park}, \citenamefont
  {Durairaj}, \citenamefont {Cao},\ and\ \citenamefont {Rotenberg}}]{Kim08}%
  \BibitemOpen
  \bibfield  {author} {\bibinfo {author} {\bibfnamefont {B.~J.}\ \bibnamefont
  {Kim}}, \bibinfo {author} {\bibfnamefont {Hosub}\ \bibnamefont {Jin}},
  \bibinfo {author} {\bibfnamefont {S.~J.}\ \bibnamefont {Moon}}, \bibinfo
  {author} {\bibfnamefont {J.-Y.}\ \bibnamefont {Kim}}, \bibinfo {author}
  {\bibfnamefont {B.-G.}\ \bibnamefont {Park}}, \bibinfo {author}
  {\bibfnamefont {C.~S.}\ \bibnamefont {Leem}}, \bibinfo {author}
  {\bibfnamefont {Jaejun}\ \bibnamefont {Yu}}, \bibinfo {author} {\bibfnamefont
  {T.~W.}\ \bibnamefont {Noh}}, \bibinfo {author} {\bibfnamefont
  {C.}~\bibnamefont {Kim}}, \bibinfo {author} {\bibfnamefont {S.-J.}\
  \bibnamefont {Oh}}, \bibinfo {author} {\bibfnamefont {J.-H.}\ \bibnamefont
  {Park}}, \bibinfo {author} {\bibfnamefont {V.}~\bibnamefont {Durairaj}},
  \bibinfo {author} {\bibfnamefont {G.}~\bibnamefont {Cao}}, \ and\ \bibinfo
  {author} {\bibfnamefont {E.}~\bibnamefont {Rotenberg}},\ }\bibfield  {title}
  {\enquote {\bibinfo {title} {{Novel ${J}_\mathrm{eff}=1/2$ Mott State Induced
  by Relativistic Spin-Orbit Coupling in $\mathrm{Sr_2IrO_4}$}},}\ }\href
  {\doibase 10.1103/PhysRevLett.101.076402} {\bibfield  {journal} {\bibinfo
  {journal} {Phys. Rev. Lett.}\ }\textbf {\bibinfo {volume} {101}},\ \bibinfo
  {pages} {076402} (\bibinfo {year} {2008})}\BibitemShut {NoStop}%
\bibitem [{\citenamefont {Kim}\ \emph {et~al.}(2009)\citenamefont {Kim},
  \citenamefont {Ohsumi}, \citenamefont {Komesu}, \citenamefont {Sakai},
  \citenamefont {Morita}, \citenamefont {Takagi},\ and\ \citenamefont
  {Arima}}]{Kim09}%
  \BibitemOpen
  \bibfield  {author} {\bibinfo {author} {\bibfnamefont {B.~J.}\ \bibnamefont
  {Kim}}, \bibinfo {author} {\bibfnamefont {H.}~\bibnamefont {Ohsumi}},
  \bibinfo {author} {\bibfnamefont {T.}~\bibnamefont {Komesu}}, \bibinfo
  {author} {\bibfnamefont {S.}~\bibnamefont {Sakai}}, \bibinfo {author}
  {\bibfnamefont {T.}~\bibnamefont {Morita}}, \bibinfo {author} {\bibfnamefont
  {H.}~\bibnamefont {Takagi}}, \ and\ \bibinfo {author} {\bibfnamefont
  {T.}~\bibnamefont {Arima}},\ }\bibfield  {title} {\enquote {\bibinfo {title}
  {{Phase-Sensitive Observation of a Spin-Orbital Mott State in
  $\mathrm{Sr_2IrO_4}$}},}\ }\href {\doibase 10.1126/science.1167106}
  {\bibfield  {journal} {\bibinfo  {journal} {Science}\ }\textbf {\bibinfo
  {volume} {323}},\ \bibinfo {pages} {1329–1332} (\bibinfo {year}
  {2009})}\BibitemShut {NoStop}%
\bibitem [{\citenamefont {Jackeli}\ and\ \citenamefont
  {Khaliullin}(2009)}]{Jac09}%
  \BibitemOpen
  \bibfield  {author} {\bibinfo {author} {\bibfnamefont {G.}~\bibnamefont
  {Jackeli}}\ and\ \bibinfo {author} {\bibfnamefont {G.}~\bibnamefont
  {Khaliullin}},\ }\bibfield  {title} {\enquote {\bibinfo {title} {{Mott
  Insulators in the Strong Spin-Orbit Coupling Limit: From Heisenberg to a
  Quantum Compass and Kitaev Models}},}\ }\href {\doibase
  10.1103/PhysRevLett.102.017205} {\bibfield  {journal} {\bibinfo  {journal}
  {Phys. Rev. Lett.}\ }\textbf {\bibinfo {volume} {102}},\ \bibinfo {pages}
  {017205} (\bibinfo {year} {2009})}\BibitemShut {NoStop}%
\bibitem [{\citenamefont {Bertinshaw}\ \emph {et~al.}(2019)\citenamefont
  {Bertinshaw}, \citenamefont {Kim}, \citenamefont {Khaliullin},\ and\
  \citenamefont {Kim}}]{Ber19}%
  \BibitemOpen
  \bibfield  {author} {\bibinfo {author} {\bibfnamefont {Joel}\ \bibnamefont
  {Bertinshaw}}, \bibinfo {author} {\bibfnamefont {Y.~K.}\ \bibnamefont {Kim}},
  \bibinfo {author} {\bibfnamefont {Giniyat}\ \bibnamefont {Khaliullin}}, \
  and\ \bibinfo {author} {\bibfnamefont {B.~J.}\ \bibnamefont {Kim}},\
  }\bibfield  {title} {\enquote {\bibinfo {title} {{Square Lattice
  Iridates}},}\ }\href {\doibase 10.1146/annurev-conmatphys-031218-013113}
  {\bibfield  {journal} {\bibinfo  {journal} {Annu. Rev. Condens. Matter
  Phys.}\ }\textbf {\bibinfo {volume} {10}},\ \bibinfo {pages} {315–336}
  (\bibinfo {year} {2019})}\BibitemShut {NoStop}%
\bibitem [{\citenamefont {Dalla~Piazza}\ \emph {et~al.}(2015)\citenamefont
  {Dalla~Piazza}, \citenamefont {Mourigal}, \citenamefont {Christensen},
  \citenamefont {Nilsen}, \citenamefont {Tregenna-Piggott}, \citenamefont
  {Perring}, \citenamefont {Enderle}, \citenamefont {McMorrow}, \citenamefont
  {Ivanov},\ and\ \citenamefont {R{\o}nnow}}]{Pia15}%
  \BibitemOpen
  \bibfield  {author} {\bibinfo {author} {\bibfnamefont {B.}~\bibnamefont
  {Dalla~Piazza}}, \bibinfo {author} {\bibfnamefont {M.}~\bibnamefont
  {Mourigal}}, \bibinfo {author} {\bibfnamefont {N.~B.}\ \bibnamefont
  {Christensen}}, \bibinfo {author} {\bibfnamefont {G.~J.}\ \bibnamefont
  {Nilsen}}, \bibinfo {author} {\bibfnamefont {P.}~\bibnamefont
  {Tregenna-Piggott}}, \bibinfo {author} {\bibfnamefont {T.~G.}\ \bibnamefont
  {Perring}}, \bibinfo {author} {\bibfnamefont {M.}~\bibnamefont {Enderle}},
  \bibinfo {author} {\bibfnamefont {D.~F.}\ \bibnamefont {McMorrow}}, \bibinfo
  {author} {\bibfnamefont {D.~A.}\ \bibnamefont {Ivanov}}, \ and\ \bibinfo
  {author} {\bibfnamefont {H.~M.}\ \bibnamefont {R{\o}nnow}},\ }\bibfield
  {title} {\enquote {\bibinfo {title} {{Fractional Excitations in the
  Square-Lattice Quantum Antiferromagnet}},}\ }\href {\doibase
  10.1038/nphys3172} {\bibfield  {journal} {\bibinfo  {journal} {Nat. Phys.}\
  }\textbf {\bibinfo {volume} {11}},\ \bibinfo {pages} {62–68} (\bibinfo
  {year} {2015})}\BibitemShut {NoStop}%
\bibitem [{\citenamefont {Shao}\ \emph {et~al.}(2017)\citenamefont {Shao},
  \citenamefont {Qin}, \citenamefont {Capponi}, \citenamefont {Chesi},
  \citenamefont {Meng},\ and\ \citenamefont {Sandvik}}]{Sha17}%
  \BibitemOpen
  \bibfield  {author} {\bibinfo {author} {\bibfnamefont {Hui}\ \bibnamefont
  {Shao}}, \bibinfo {author} {\bibfnamefont {Yan~Qi}\ \bibnamefont {Qin}},
  \bibinfo {author} {\bibfnamefont {Sylvain}\ \bibnamefont {Capponi}}, \bibinfo
  {author} {\bibfnamefont {Stefano}\ \bibnamefont {Chesi}}, \bibinfo {author}
  {\bibfnamefont {Zi~Yang}\ \bibnamefont {Meng}}, \ and\ \bibinfo {author}
  {\bibfnamefont {Anders~W.}\ \bibnamefont {Sandvik}},\ }\bibfield  {title}
  {\enquote {\bibinfo {title} {{Nearly Deconfined Spinon Excitations in the
  Square-Lattice Spin-$1/2$ Heisenberg Antiferromagnet}},}\ }\href {\doibase
  10.1103/PhysRevX.7.041072} {\bibfield  {journal} {\bibinfo  {journal} {Phys.
  Rev. X}\ }\textbf {\bibinfo {volume} {7}},\ \bibinfo {pages} {041072}
  (\bibinfo {year} {2017})}\BibitemShut {NoStop}%
\bibitem [{\citenamefont {Anderson}(1987)}]{And87}%
  \BibitemOpen
  \bibfield  {author} {\bibinfo {author} {\bibfnamefont {P.~W.}\ \bibnamefont
  {Anderson}},\ }\bibfield  {title} {\enquote {\bibinfo {title} {{The
  Resonating Valence Bond State in $\mathrm{La}_{2}\mathrm{CuO}_{4}$ and
  Superconductivity}},}\ }\href {\doibase 10.1126/science.235.4793.1196}
  {\bibfield  {journal} {\bibinfo  {journal} {Science}\ }\textbf {\bibinfo
  {volume} {235}},\ \bibinfo {pages} {1196–1198} (\bibinfo {year}
  {1987})}\BibitemShut {NoStop}%
\bibitem [{\citenamefont {Lee}\ \emph {et~al.}(2006)\citenamefont {Lee},
  \citenamefont {Nagaosa},\ and\ \citenamefont {Wen}}]{Pat06}%
  \BibitemOpen
  \bibfield  {author} {\bibinfo {author} {\bibfnamefont {Patrick~A.}\
  \bibnamefont {Lee}}, \bibinfo {author} {\bibfnamefont {Naoto}\ \bibnamefont
  {Nagaosa}}, \ and\ \bibinfo {author} {\bibfnamefont {Xiao-Gang}\ \bibnamefont
  {Wen}},\ }\bibfield  {title} {\enquote {\bibinfo {title} {{Doping a Mott
  Insulator: Physics of High-temperature Superconductivity}},}\ }\href
  {\doibase 10.1103/RevModPhys.78.17} {\bibfield  {journal} {\bibinfo
  {journal} {Rev. Mod. Phys.}\ }\textbf {\bibinfo {volume} {78}},\ \bibinfo
  {pages} {17–85} (\bibinfo {year} {2006})}\BibitemShut {NoStop}%
\bibitem [{\citenamefont {Keimer}\ \emph {et~al.}(2015)\citenamefont {Keimer},
  \citenamefont {Kivelson}, \citenamefont {Norman}, \citenamefont {Uchida},\
  and\ \citenamefont {Zaanen}}]{Kei15}%
  \BibitemOpen
  \bibfield  {author} {\bibinfo {author} {\bibfnamefont {B.}~\bibnamefont
  {Keimer}}, \bibinfo {author} {\bibfnamefont {S.~A.}\ \bibnamefont
  {Kivelson}}, \bibinfo {author} {\bibfnamefont {M.~R.}\ \bibnamefont
  {Norman}}, \bibinfo {author} {\bibfnamefont {S.}~\bibnamefont {Uchida}}, \
  and\ \bibinfo {author} {\bibfnamefont {J.}~\bibnamefont {Zaanen}},\
  }\bibfield  {title} {\enquote {\bibinfo {title} {{From quantum matter to
  high-temperature superconductivity in copper oxides}},}\ }\href {\doibase
  10.1038/nature14165} {\bibfield  {journal} {\bibinfo  {journal} {Nature}\
  }\textbf {\bibinfo {volume} {518}},\ \bibinfo {pages} {179–186} (\bibinfo
  {year} {2015})}\BibitemShut {NoStop}%
\bibitem [{\citenamefont {Scalapino}(2012)}]{Sca12}%
  \BibitemOpen
  \bibfield  {author} {\bibinfo {author} {\bibfnamefont {D.~J.}\ \bibnamefont
  {Scalapino}},\ }\bibfield  {title} {\enquote {\bibinfo {title} {{A common
  thread: The pairing interaction for unconventional superconductors}},}\
  }\href {\doibase 10.1103/RevModPhys.84.1383} {\bibfield  {journal} {\bibinfo
  {journal} {Rev. Mod. Phys.}\ }\textbf {\bibinfo {volume} {84}},\ \bibinfo
  {pages} {1383–1417} (\bibinfo {year} {2012})}\BibitemShut {NoStop}%
\bibitem [{\citenamefont {Vaknin}\ \emph {et~al.}(1987)\citenamefont {Vaknin},
  \citenamefont {Sinha}, \citenamefont {Moncton}, \citenamefont {Johnston},
  \citenamefont {Newsam}, \citenamefont {Safinya},\ and\ \citenamefont
  {King}}]{Vak87}%
  \BibitemOpen
  \bibfield  {author} {\bibinfo {author} {\bibfnamefont {D.}~\bibnamefont
  {Vaknin}}, \bibinfo {author} {\bibfnamefont {S.~K.}\ \bibnamefont {Sinha}},
  \bibinfo {author} {\bibfnamefont {D.~E.}\ \bibnamefont {Moncton}}, \bibinfo
  {author} {\bibfnamefont {D.~C.}\ \bibnamefont {Johnston}}, \bibinfo {author}
  {\bibfnamefont {J.~M.}\ \bibnamefont {Newsam}}, \bibinfo {author}
  {\bibfnamefont {C.~R.}\ \bibnamefont {Safinya}}, \ and\ \bibinfo {author}
  {\bibfnamefont {H.~E.}\ \bibnamefont {King}},\ }\bibfield  {title} {\enquote
  {\bibinfo {title} {{Antiferromagnetism in
  ${\mathrm{La}}_{2}$${\mathrm{CuO}}_{4\mathrm{\ensuremath{-}}\mathrm{y}}$}},}\
  }\href {\doibase 10.1103/PhysRevLett.58.2802} {\bibfield  {journal} {\bibinfo
   {journal} {Phys. Rev. Lett.}\ }\textbf {\bibinfo {volume} {58}},\ \bibinfo
  {pages} {2802–2805} (\bibinfo {year} {1987})}\BibitemShut {NoStop}%
\bibitem [{\citenamefont {Singh}(1989)}]{Sin89}%
  \BibitemOpen
  \bibfield  {author} {\bibinfo {author} {\bibfnamefont {Rajiv R.~P.}\
  \bibnamefont {Singh}},\ }\bibfield  {title} {\enquote {\bibinfo {title}
  {{Thermodynamic parameters of the T=0, spin-1/2 square-lattice Heisenberg
  antiferromagnet}},}\ }\href {\doibase 10.1103/PhysRevB.39.9760} {\bibfield
  {journal} {\bibinfo  {journal} {Phys. Rev. B}\ }\textbf {\bibinfo {volume}
  {39}},\ \bibinfo {pages} {9760–9763} (\bibinfo {year} {1989})}\BibitemShut
  {NoStop}%
\bibitem [{\citenamefont {Coldea}\ \emph {et~al.}(2001)\citenamefont {Coldea},
  \citenamefont {Hayden}, \citenamefont {Aeppli}, \citenamefont {Perring},
  \citenamefont {Frost}, \citenamefont {Mason}, \citenamefont {Cheong},\ and\
  \citenamefont {Fisk}}]{Col01}%
  \BibitemOpen
  \bibfield  {author} {\bibinfo {author} {\bibfnamefont {R.}~\bibnamefont
  {Coldea}}, \bibinfo {author} {\bibfnamefont {S.~M.}\ \bibnamefont {Hayden}},
  \bibinfo {author} {\bibfnamefont {G.}~\bibnamefont {Aeppli}}, \bibinfo
  {author} {\bibfnamefont {T.~G.}\ \bibnamefont {Perring}}, \bibinfo {author}
  {\bibfnamefont {C.~D.}\ \bibnamefont {Frost}}, \bibinfo {author}
  {\bibfnamefont {T.~E.}\ \bibnamefont {Mason}}, \bibinfo {author}
  {\bibfnamefont {S.-W.}\ \bibnamefont {Cheong}}, \ and\ \bibinfo {author}
  {\bibfnamefont {Z.}~\bibnamefont {Fisk}},\ }\bibfield  {title} {\enquote
  {\bibinfo {title} {{Spin Waves and Electronic Interactions in
  ${\mathrm{La}}_{2}{\mathrm{CuO}}_{4}$}},}\ }\href {\doibase
  10.1103/PhysRevLett.86.5377} {\bibfield  {journal} {\bibinfo  {journal}
  {Phys. Rev. Lett.}\ }\textbf {\bibinfo {volume} {86}},\ \bibinfo {pages}
  {5377–5380} (\bibinfo {year} {2001})}\BibitemShut {NoStop}%
\bibitem [{\citenamefont {Savary}\ and\ \citenamefont {Balents}(2016)}]{Sav16}%
  \BibitemOpen
  \bibfield  {author} {\bibinfo {author} {\bibfnamefont {Lucile}\ \bibnamefont
  {Savary}}\ and\ \bibinfo {author} {\bibfnamefont {Leon}\ \bibnamefont
  {Balents}},\ }\bibfield  {title} {\enquote {\bibinfo {title} {{Quantum Spin
  Liquids: a Review}},}\ }\href {\doibase 10.1088/0034-4885/80/1/016502}
  {\bibfield  {journal} {\bibinfo  {journal} {Rep. Prog. Phys.}\ }\textbf
  {\bibinfo {volume} {80}},\ \bibinfo {pages} {016502} (\bibinfo {year}
  {2016})}\BibitemShut {NoStop}%
\bibitem [{\citenamefont {Zhou}\ \emph {et~al.}(2017)\citenamefont {Zhou},
  \citenamefont {Kanoda},\ and\ \citenamefont {Ng}}]{Zho17}%
  \BibitemOpen
  \bibfield  {author} {\bibinfo {author} {\bibfnamefont {Yi}~\bibnamefont
  {Zhou}}, \bibinfo {author} {\bibfnamefont {Kazushi}\ \bibnamefont {Kanoda}},
  \ and\ \bibinfo {author} {\bibfnamefont {Tai-Kai}\ \bibnamefont {Ng}},\
  }\bibfield  {title} {\enquote {\bibinfo {title} {{Quantum Spin Liquid
  States}},}\ }\href {\doibase 10.1103/RevModPhys.89.025003} {\bibfield
  {journal} {\bibinfo  {journal} {Rev. Mod. Phys.}\ }\textbf {\bibinfo {volume}
  {89}},\ \bibinfo {pages} {025003} (\bibinfo {year} {2017})}\BibitemShut
  {NoStop}%
\bibitem [{\citenamefont {Barzykin}\ and\ \citenamefont
  {Gor'kov}(1993)}]{Bar93}%
  \BibitemOpen
  \bibfield  {author} {\bibinfo {author} {\bibfnamefont {V.}~\bibnamefont
  {Barzykin}}\ and\ \bibinfo {author} {\bibfnamefont {L.~P.}\ \bibnamefont
  {Gor'kov}},\ }\bibfield  {title} {\enquote {\bibinfo {title} {{Possibility of
  Observation of Nontrivial Magnetic Order by Elastic Neutron Scattering in
  Magnetic Field}},}\ }\href {\doibase 10.1103/PhysRevLett.70.2479} {\bibfield
  {journal} {\bibinfo  {journal} {Phys. Rev. Lett.}\ }\textbf {\bibinfo
  {volume} {70}},\ \bibinfo {pages} {2479–2482} (\bibinfo {year}
  {1993})}\BibitemShut {NoStop}%
\bibitem [{\citenamefont {Smerald}\ and\ \citenamefont
  {Shannon}(2013)}]{Sme13}%
  \BibitemOpen
  \bibfield  {author} {\bibinfo {author} {\bibfnamefont {Andrew}\ \bibnamefont
  {Smerald}}\ and\ \bibinfo {author} {\bibfnamefont {Nic}\ \bibnamefont
  {Shannon}},\ }\bibfield  {title} {\enquote {\bibinfo {title} {{Theory of Spin
  Excitations in a Quantum Spin-nematic State}},}\ }\href {\doibase
  10.1103/PhysRevB.88.184430} {\bibfield  {journal} {\bibinfo  {journal} {Phys.
  Rev. B}\ }\textbf {\bibinfo {volume} {88}},\ \bibinfo {pages} {184430}
  (\bibinfo {year} {2013})}\BibitemShut {NoStop}%
\bibitem [{\citenamefont {Savary}\ and\ \citenamefont {Senthil}(2015)}]{Sav15}%
  \BibitemOpen
  \bibfield  {author} {\bibinfo {author} {\bibfnamefont {Lucile}\ \bibnamefont
  {Savary}}\ and\ \bibinfo {author} {\bibfnamefont {T.}~\bibnamefont
  {Senthil}},\ }\bibfield  {title} {\enquote {\bibinfo {title} {{Probing Hidden
  Orders with Resonant Inelastic X-Ray Scattering}},}\ }\href
  {https://arxiv.org/abs/1506.04752} {\bibfield  {journal} {\bibinfo  {journal}
  {arXiv:1506.04752}\ } (\bibinfo {year} {2015})}\BibitemShut {NoStop}%
\bibitem [{\citenamefont {Kohama}\ \emph {et~al.}(2019)\citenamefont {Kohama},
  \citenamefont {Ishikawa}, \citenamefont {Matsuo}, \citenamefont {Kindo},
  \citenamefont {Shannon},\ and\ \citenamefont {Hiroi}}]{Koh19}%
  \BibitemOpen
  \bibfield  {author} {\bibinfo {author} {\bibfnamefont {Yoshimitsu}\
  \bibnamefont {Kohama}}, \bibinfo {author} {\bibfnamefont {Hajime}\
  \bibnamefont {Ishikawa}}, \bibinfo {author} {\bibfnamefont {Akira}\
  \bibnamefont {Matsuo}}, \bibinfo {author} {\bibfnamefont {Koichi}\
  \bibnamefont {Kindo}}, \bibinfo {author} {\bibfnamefont {Nic}\ \bibnamefont
  {Shannon}}, \ and\ \bibinfo {author} {\bibfnamefont {Zenji}\ \bibnamefont
  {Hiroi}},\ }\bibfield  {title} {\enquote {\bibinfo {title} {{Possible
  Observation of Quantum Spin-nematic Phase in a Frustrated Magnet}},}\ }\href
  {\doibase doi:10.1073/pnas.1821969116} {\bibfield  {journal} {\bibinfo
  {journal} {Proc. Natl. Acad. Sci. U.S.A.}\ }\textbf {\bibinfo {volume}
  {116}},\ \bibinfo {pages} {10686–10690} (\bibinfo {year}
  {2019})}\BibitemShut {NoStop}%
\bibitem [{\citenamefont {Orlova}\ \emph {et~al.}(2017)\citenamefont {Orlova},
  \citenamefont {Green}, \citenamefont {Law}, \citenamefont {Gorbunov},
  \citenamefont {Chanda}, \citenamefont {Kr\"amer}, \citenamefont
  {Horvati\ifmmode~\acute{c}\else \'{c}\fi{}}, \citenamefont {Kremer},
  \citenamefont {Wosnitza},\ and\ \citenamefont {Rikken}}]{Orl17}%
  \BibitemOpen
  \bibfield  {author} {\bibinfo {author} {\bibfnamefont {A.}~\bibnamefont
  {Orlova}}, \bibinfo {author} {\bibfnamefont {E.~L.}\ \bibnamefont {Green}},
  \bibinfo {author} {\bibfnamefont {J.~M.}\ \bibnamefont {Law}}, \bibinfo
  {author} {\bibfnamefont {D.~I.}\ \bibnamefont {Gorbunov}}, \bibinfo {author}
  {\bibfnamefont {G.}~\bibnamefont {Chanda}}, \bibinfo {author} {\bibfnamefont
  {S.}~\bibnamefont {Kr\"amer}}, \bibinfo {author} {\bibfnamefont
  {M.}~\bibnamefont {Horvati\ifmmode~\acute{c}\else \'{c}\fi{}}}, \bibinfo
  {author} {\bibfnamefont {R.~K.}\ \bibnamefont {Kremer}}, \bibinfo {author}
  {\bibfnamefont {J.}~\bibnamefont {Wosnitza}}, \ and\ \bibinfo {author}
  {\bibfnamefont {G.~L. J.~A.}\ \bibnamefont {Rikken}},\ }\bibfield  {title}
  {\enquote {\bibinfo {title} {{Nuclear Magnetic Resonance Signature of the
  Spin-Nematic Phase in ${\mathrm{LiCuVO}}_{4}$ at High Magnetic Fields}},}\
  }\href {\doibase 10.1103/PhysRevLett.118.247201} {\bibfield  {journal}
  {\bibinfo  {journal} {Phys. Rev. Lett.}\ }\textbf {\bibinfo {volume} {118}},\
  \bibinfo {pages} {247201} (\bibinfo {year} {2017})}\BibitemShut {NoStop}%
\bibitem [{\citenamefont {Ye}\ \emph {et~al.}(2013)\citenamefont {Ye},
  \citenamefont {Chi}, \citenamefont {Chakoumakos}, \citenamefont
  {Fernandez-Baca}, \citenamefont {Qi},\ and\ \citenamefont {Cao}}]{Ye13}%
  \BibitemOpen
  \bibfield  {author} {\bibinfo {author} {\bibfnamefont {Feng}\ \bibnamefont
  {Ye}}, \bibinfo {author} {\bibfnamefont {Songxue}\ \bibnamefont {Chi}},
  \bibinfo {author} {\bibfnamefont {Bryan~C.}\ \bibnamefont {Chakoumakos}},
  \bibinfo {author} {\bibfnamefont {Jaime~A.}\ \bibnamefont {Fernandez-Baca}},
  \bibinfo {author} {\bibfnamefont {Tongfei}\ \bibnamefont {Qi}}, \ and\
  \bibinfo {author} {\bibfnamefont {G.}~\bibnamefont {Cao}},\ }\bibfield
  {title} {\enquote {\bibinfo {title} {{Magnetic and Crystal Structures of
  ${\mathrm{Sr}}_{2}{\mathrm{IrO}}_{4}$: A Neutron Diffraction Study}},}\
  }\href {\doibase 10.1103/PhysRevB.87.140406} {\bibfield  {journal} {\bibinfo
  {journal} {Phys. Rev. B}\ }\textbf {\bibinfo {volume} {87}},\ \bibinfo
  {pages} {140406} (\bibinfo {year} {2013})}\BibitemShut {NoStop}%
\bibitem [{\citenamefont {Porras}\ \emph {et~al.}(2019)\citenamefont {Porras},
  \citenamefont {Bertinshaw}, \citenamefont {Liu}, \citenamefont {Khaliullin},
  \citenamefont {Sung}, \citenamefont {Kim}, \citenamefont {Francoual},
  \citenamefont {Steffens}, \citenamefont {Deng}, \citenamefont {Sala},
  \citenamefont {Efimenko}, \citenamefont {Said}, \citenamefont {Casa},
  \citenamefont {Huang}, \citenamefont {Gog}, \citenamefont {Kim},
  \citenamefont {Keimer},\ and\ \citenamefont {Kim}}]{Por19}%
  \BibitemOpen
  \bibfield  {author} {\bibinfo {author} {\bibfnamefont {J.}~\bibnamefont
  {Porras}}, \bibinfo {author} {\bibfnamefont {J.}~\bibnamefont {Bertinshaw}},
  \bibinfo {author} {\bibfnamefont {H.}~\bibnamefont {Liu}}, \bibinfo {author}
  {\bibfnamefont {G.}~\bibnamefont {Khaliullin}}, \bibinfo {author}
  {\bibfnamefont {N.~H.}\ \bibnamefont {Sung}}, \bibinfo {author}
  {\bibfnamefont {J.~W.}\ \bibnamefont {Kim}}, \bibinfo {author} {\bibfnamefont
  {S.}~\bibnamefont {Francoual}}, \bibinfo {author} {\bibfnamefont
  {P.}~\bibnamefont {Steffens}}, \bibinfo {author} {\bibfnamefont
  {G.}~\bibnamefont {Deng}}, \bibinfo {author} {\bibfnamefont {M.~Moretti}\
  \bibnamefont {Sala}}, \bibinfo {author} {\bibfnamefont {A.}~\bibnamefont
  {Efimenko}}, \bibinfo {author} {\bibfnamefont {A.}~\bibnamefont {Said}},
  \bibinfo {author} {\bibfnamefont {D.}~\bibnamefont {Casa}}, \bibinfo {author}
  {\bibfnamefont {X.}~\bibnamefont {Huang}}, \bibinfo {author} {\bibfnamefont
  {T.}~\bibnamefont {Gog}}, \bibinfo {author} {\bibfnamefont {J.}~\bibnamefont
  {Kim}}, \bibinfo {author} {\bibfnamefont {B.}~\bibnamefont {Keimer}}, \ and\
  \bibinfo {author} {\bibfnamefont {B.~J.}\ \bibnamefont {Kim}},\ }\bibfield
  {title} {\enquote {\bibinfo {title} {{Pseudospin-Lattice Coupling in the
  Spin-Orbit Mott Insulator $\mathrm{Sr_2IrO_4}$}},}\ }\href {\doibase
  10.1103/PhysRevB.99.085125} {\bibfield  {journal} {\bibinfo  {journal} {Phys.
  Rev. B}\ }\textbf {\bibinfo {volume} {99}},\ \bibinfo {pages} {085125}
  (\bibinfo {year} {2019})}\BibitemShut {NoStop}%
\bibitem [{\citenamefont {K\"onig}\ \emph {et~al.}(2022)\citenamefont
  {K\"onig}, \citenamefont {Komijani},\ and\ \citenamefont {Coleman}}]{Kon22}%
  \BibitemOpen
  \bibfield  {author} {\bibinfo {author} {\bibfnamefont {Elio~J.}\ \bibnamefont
  {K\"onig}}, \bibinfo {author} {\bibfnamefont {Yashar}\ \bibnamefont
  {Komijani}}, \ and\ \bibinfo {author} {\bibfnamefont {Piers}\ \bibnamefont
  {Coleman}},\ }\bibfield  {title} {\enquote {\bibinfo {title} {{Triplet
  Resonating Valence Bond Theory and Transition Metal Chalcogenides}},}\ }\href
  {\doibase 10.1103/PhysRevB.105.075142} {\bibfield  {journal} {\bibinfo
  {journal} {Phys. Rev. B}\ }\textbf {\bibinfo {volume} {105}},\ \bibinfo
  {pages} {075142} (\bibinfo {year} {2022})}\BibitemShut {NoStop}%
\bibitem [{\citenamefont {Shindou}\ and\ \citenamefont {Momoi}(2009)}]{Shi09}%
  \BibitemOpen
  \bibfield  {author} {\bibinfo {author} {\bibfnamefont {Ryuichi}\ \bibnamefont
  {Shindou}}\ and\ \bibinfo {author} {\bibfnamefont {Tsutomu}\ \bibnamefont
  {Momoi}},\ }\bibfield  {title} {\enquote {\bibinfo {title} {{$SU(2)$
  Slave-boson Formulation of Spin Nematic States in $S=\frac{1}{2}$ Frustrated
  Ferromagnets}},}\ }\href {\doibase 10.1103/PhysRevB.80.064410} {\bibfield
  {journal} {\bibinfo  {journal} {Phys. Rev. B}\ }\textbf {\bibinfo {volume}
  {80}},\ \bibinfo {pages} {064410} (\bibinfo {year} {2009})}\BibitemShut
  {NoStop}%
\bibitem [{\citenamefont {Cetin}\ \emph {et~al.}(2012)\citenamefont {Cetin},
  \citenamefont {Lemmens}, \citenamefont {Gnezdilov}, \citenamefont
  {Wulferding}, \citenamefont {Menzel}, \citenamefont {Takayama}, \citenamefont
  {Ohashi},\ and\ \citenamefont {Takagi}}]{Cet12}%
  \BibitemOpen
  \bibfield  {author} {\bibinfo {author} {\bibfnamefont {Mehmet~Fatih}\
  \bibnamefont {Cetin}}, \bibinfo {author} {\bibfnamefont {Peter}\ \bibnamefont
  {Lemmens}}, \bibinfo {author} {\bibfnamefont {Vladimir}\ \bibnamefont
  {Gnezdilov}}, \bibinfo {author} {\bibfnamefont {Dirk}\ \bibnamefont
  {Wulferding}}, \bibinfo {author} {\bibfnamefont {Dirk}\ \bibnamefont
  {Menzel}}, \bibinfo {author} {\bibfnamefont {Tomohiro}\ \bibnamefont
  {Takayama}}, \bibinfo {author} {\bibfnamefont {Kei}\ \bibnamefont {Ohashi}},
  \ and\ \bibinfo {author} {\bibfnamefont {Hidenori}\ \bibnamefont {Takagi}},\
  }\bibfield  {title} {\enquote {\bibinfo {title} {{Crossover from Coherent to
  Incoherent Scattering in Spin-Orbit Dominated
  ${\mathrm{Sr}}_{2}{\mathrm{IrO}}_{4}$}},}\ }\href {\doibase
  10.1103/PhysRevB.85.195148} {\bibfield  {journal} {\bibinfo  {journal} {Phys.
  Rev. B}\ }\textbf {\bibinfo {volume} {85}},\ \bibinfo {pages} {195148}
  (\bibinfo {year} {2012})}\BibitemShut {NoStop}%
\bibitem [{\citenamefont {Gim}\ \emph {et~al.}(2016)\citenamefont {Gim},
  \citenamefont {Sethi}, \citenamefont {Zhao}, \citenamefont {Mitchell},
  \citenamefont {Cao},\ and\ \citenamefont {Cooper}}]{Gim16}%
  \BibitemOpen
  \bibfield  {author} {\bibinfo {author} {\bibfnamefont {Y.}~\bibnamefont
  {Gim}}, \bibinfo {author} {\bibfnamefont {A.}~\bibnamefont {Sethi}}, \bibinfo
  {author} {\bibfnamefont {Q.}~\bibnamefont {Zhao}}, \bibinfo {author}
  {\bibfnamefont {J.~F.}\ \bibnamefont {Mitchell}}, \bibinfo {author}
  {\bibfnamefont {G.}~\bibnamefont {Cao}}, \ and\ \bibinfo {author}
  {\bibfnamefont {S.~L.}\ \bibnamefont {Cooper}},\ }\bibfield  {title}
  {\enquote {\bibinfo {title} {{Isotropic and Anisotropic Regimes of the
  Field-dependent Spin Dynamics in ${\mathrm{Sr}}_{2}{\mathrm{IrO}}_{4}$: Raman
  Scattering Studies}},}\ }\href {\doibase 10.1103/PhysRevB.93.024405}
  {\bibfield  {journal} {\bibinfo  {journal} {Phys. Rev. B}\ }\textbf {\bibinfo
  {volume} {93}},\ \bibinfo {pages} {024405} (\bibinfo {year}
  {2016})}\BibitemShut {NoStop}%
\bibitem [{\citenamefont {Gretarsson}\ \emph {et~al.}(2016)\citenamefont
  {Gretarsson}, \citenamefont {Sung}, \citenamefont {H\"oppner}, \citenamefont
  {Kim}, \citenamefont {Keimer},\ and\ \citenamefont {Le~Tacon}}]{Gre16}%
  \BibitemOpen
  \bibfield  {author} {\bibinfo {author} {\bibfnamefont {H.}~\bibnamefont
  {Gretarsson}}, \bibinfo {author} {\bibfnamefont {N.~H.}\ \bibnamefont
  {Sung}}, \bibinfo {author} {\bibfnamefont {M.}~\bibnamefont {H\"oppner}},
  \bibinfo {author} {\bibfnamefont {B.~J.}\ \bibnamefont {Kim}}, \bibinfo
  {author} {\bibfnamefont {B.}~\bibnamefont {Keimer}}, \ and\ \bibinfo {author}
  {\bibfnamefont {M.}~\bibnamefont {Le~Tacon}},\ }\bibfield  {title} {\enquote
  {\bibinfo {title} {{Two-Magnon Raman Scattering and Pseudospin-Lattice
  Interactions in ${\mathrm{Sr}}_{2}{\mathrm{IrO}}_{4}$ and
  ${\mathrm{Sr}}_{3}{\mathrm{Ir}}_{2}{\mathrm{O}}_{7}$}},}\ }\href {\doibase
  10.1103/PhysRevLett.116.136401} {\bibfield  {journal} {\bibinfo  {journal}
  {Phys. Rev. Lett.}\ }\textbf {\bibinfo {volume} {116}},\ \bibinfo {pages}
  {136401} (\bibinfo {year} {2016})}\BibitemShut {NoStop}%
\bibitem [{\citenamefont {Aeppli}\ \emph {et~al.}(1989)\citenamefont {Aeppli},
  \citenamefont {Hayden}, \citenamefont {Mook}, \citenamefont {Fisk},
  \citenamefont {Cheong}, \citenamefont {Rytz}, \citenamefont {Remeika},
  \citenamefont {Espinosa},\ and\ \citenamefont {Cooper}}]{Aep89}%
  \BibitemOpen
  \bibfield  {author} {\bibinfo {author} {\bibfnamefont {G.}~\bibnamefont
  {Aeppli}}, \bibinfo {author} {\bibfnamefont {S.~M.}\ \bibnamefont {Hayden}},
  \bibinfo {author} {\bibfnamefont {H.~A.}\ \bibnamefont {Mook}}, \bibinfo
  {author} {\bibfnamefont {Z.}~\bibnamefont {Fisk}}, \bibinfo {author}
  {\bibfnamefont {S.-W.}\ \bibnamefont {Cheong}}, \bibinfo {author}
  {\bibfnamefont {D.}~\bibnamefont {Rytz}}, \bibinfo {author} {\bibfnamefont
  {J.~P.}\ \bibnamefont {Remeika}}, \bibinfo {author} {\bibfnamefont {G.~P.}\
  \bibnamefont {Espinosa}}, \ and\ \bibinfo {author} {\bibfnamefont {A.~S.}\
  \bibnamefont {Cooper}},\ }\bibfield  {title} {\enquote {\bibinfo {title}
  {{Magnetic dynamics of ${\mathrm{La}}_{2}$${\mathrm{CuO}}_{4}$ and
  ${\mathrm{La}}_{2\mathrm{\ensuremath{-}}\mathrm{x}}$${\mathrm{Ba}}_{\mathrm{x}}$${\mathrm{CuO}}_{4}$}},}\
  }\href {\doibase 10.1103/PhysRevLett.62.2052} {\bibfield  {journal} {\bibinfo
   {journal} {Phys. Rev. Lett.}\ }\textbf {\bibinfo {volume} {62}},\ \bibinfo
  {pages} {2052–2055} (\bibinfo {year} {1989})}\BibitemShut {NoStop}%
\bibitem [{\citenamefont {Braicovich}\ \emph {et~al.}(2014)\citenamefont
  {Braicovich}, \citenamefont {Minola}, \citenamefont {Dellea}, \citenamefont
  {Le~Tacon}, \citenamefont {Moretti~Sala}, \citenamefont {Morawe},
  \citenamefont {Peffen}, \citenamefont {Supruangnet}, \citenamefont {Yakhou},
  \citenamefont {Ghiringhelli},\ and\ \citenamefont {Brookes}}]{Bra14}%
  \BibitemOpen
  \bibfield  {author} {\bibinfo {author} {\bibfnamefont {L.}~\bibnamefont
  {Braicovich}}, \bibinfo {author} {\bibfnamefont {M.}~\bibnamefont {Minola}},
  \bibinfo {author} {\bibfnamefont {G.}~\bibnamefont {Dellea}}, \bibinfo
  {author} {\bibfnamefont {M.}~\bibnamefont {Le~Tacon}}, \bibinfo {author}
  {\bibfnamefont {M.}~\bibnamefont {Moretti~Sala}}, \bibinfo {author}
  {\bibfnamefont {C.}~\bibnamefont {Morawe}}, \bibinfo {author} {\bibfnamefont
  {J.-Ch.}\ \bibnamefont {Peffen}}, \bibinfo {author} {\bibfnamefont
  {R.}~\bibnamefont {Supruangnet}}, \bibinfo {author} {\bibfnamefont
  {F.}~\bibnamefont {Yakhou}}, \bibinfo {author} {\bibfnamefont
  {G.}~\bibnamefont {Ghiringhelli}}, \ and\ \bibinfo {author} {\bibfnamefont
  {N.~B.}\ \bibnamefont {Brookes}},\ }\bibfield  {title} {\enquote {\bibinfo
  {title} {{The Simultaneous Measurement of Energy and Linear Polarization of
  the Scattered Radiation in Resonant Inelastic Soft X-ray Scattering}},}\
  }\href {\doibase 10.1063/1.4900959} {\bibfield  {journal} {\bibinfo
  {journal} {Rev. Sci. Instrum.}\ }\textbf {\bibinfo {volume} {85}},\ \bibinfo
  {pages} {115104} (\bibinfo {year} {2014})}\BibitemShut {NoStop}%
\bibitem [{\citenamefont {Peng}\ \emph {et~al.}(2017)\citenamefont {Peng},
  \citenamefont {Dellea}, \citenamefont {Minola}, \citenamefont {Conni},
  \citenamefont {Amorese}, \citenamefont {Di~Castro}, \citenamefont {De~Luca},
  \citenamefont {Kummer}, \citenamefont {Salluzzo}, \citenamefont {Sun},
  \citenamefont {Zhou}, \citenamefont {Balestrino}, \citenamefont {Le~Tacon},
  \citenamefont {Keimer}, \citenamefont {Braicovich}, \citenamefont {Brookes},\
  and\ \citenamefont {Ghiringhelli}}]{Pen17}%
  \BibitemOpen
  \bibfield  {author} {\bibinfo {author} {\bibfnamefont {Y.~Y.}\ \bibnamefont
  {Peng}}, \bibinfo {author} {\bibfnamefont {G.}~\bibnamefont {Dellea}},
  \bibinfo {author} {\bibfnamefont {M.}~\bibnamefont {Minola}}, \bibinfo
  {author} {\bibfnamefont {M.}~\bibnamefont {Conni}}, \bibinfo {author}
  {\bibfnamefont {A.}~\bibnamefont {Amorese}}, \bibinfo {author} {\bibfnamefont
  {D.}~\bibnamefont {Di~Castro}}, \bibinfo {author} {\bibfnamefont {G.~M.}\
  \bibnamefont {De~Luca}}, \bibinfo {author} {\bibfnamefont {K.}~\bibnamefont
  {Kummer}}, \bibinfo {author} {\bibfnamefont {M.}~\bibnamefont {Salluzzo}},
  \bibinfo {author} {\bibfnamefont {X.}~\bibnamefont {Sun}}, \bibinfo {author}
  {\bibfnamefont {X.~J.}\ \bibnamefont {Zhou}}, \bibinfo {author}
  {\bibfnamefont {G.}~\bibnamefont {Balestrino}}, \bibinfo {author}
  {\bibfnamefont {M.}~\bibnamefont {Le~Tacon}}, \bibinfo {author}
  {\bibfnamefont {B.}~\bibnamefont {Keimer}}, \bibinfo {author} {\bibfnamefont
  {L.}~\bibnamefont {Braicovich}}, \bibinfo {author} {\bibfnamefont {N.~B.}\
  \bibnamefont {Brookes}}, \ and\ \bibinfo {author} {\bibfnamefont
  {G.}~\bibnamefont {Ghiringhelli}},\ }\bibfield  {title} {\enquote {\bibinfo
  {title} {{Influence of Apical Oxygen on the Extent of In-Plane Exchange
  Interaction in Cuprate Superconductors}},}\ }\href {\doibase
  10.1038/nphys4248} {\bibfield  {journal} {\bibinfo  {journal} {Nat. Phys.}\
  }\textbf {\bibinfo {volume} {13}},\ \bibinfo {pages} {1201–1206} (\bibinfo
  {year} {2017})}\BibitemShut {NoStop}%
\bibitem [{\citenamefont {Christensen}\ \emph {et~al.}(2007)\citenamefont
  {Christensen}, \citenamefont {R{\o}nnow}, \citenamefont {McMorrow},
  \citenamefont {Harrison}, \citenamefont {Perring}, \citenamefont {Enderle},
  \citenamefont {Coldea}, \citenamefont {Regnault},\ and\ \citenamefont
  {Aeppli}}]{Chr07}%
  \BibitemOpen
  \bibfield  {author} {\bibinfo {author} {\bibfnamefont {N.~B.}\ \bibnamefont
  {Christensen}}, \bibinfo {author} {\bibfnamefont {H.~M.}\ \bibnamefont
  {R{\o}nnow}}, \bibinfo {author} {\bibfnamefont {D.~F.}\ \bibnamefont
  {McMorrow}}, \bibinfo {author} {\bibfnamefont {A.}~\bibnamefont {Harrison}},
  \bibinfo {author} {\bibfnamefont {T.~G.}\ \bibnamefont {Perring}}, \bibinfo
  {author} {\bibfnamefont {M.}~\bibnamefont {Enderle}}, \bibinfo {author}
  {\bibfnamefont {R.}~\bibnamefont {Coldea}}, \bibinfo {author} {\bibfnamefont
  {L.~P.}\ \bibnamefont {Regnault}}, \ and\ \bibinfo {author} {\bibfnamefont
  {G.}~\bibnamefont {Aeppli}},\ }\bibfield  {title} {\enquote {\bibinfo {title}
  {{Quantum Dynamics and Entanglement of Spins on a Square Lattice}},}\ }\href
  {\doibase 10.1073/pnas.0703293104} {\bibfield  {journal} {\bibinfo  {journal}
  {Proc. Natl. Acad. Sci. U.S.A.}\ }\textbf {\bibinfo {volume} {104}},\
  \bibinfo {pages} {15264–15269} (\bibinfo {year} {2007})}\BibitemShut
  {NoStop}%
\bibitem [{\citenamefont {Headings}\ \emph {et~al.}(2010)\citenamefont
  {Headings}, \citenamefont {Hayden}, \citenamefont {Coldea},\ and\
  \citenamefont {Perring}}]{Hea10}%
  \BibitemOpen
  \bibfield  {author} {\bibinfo {author} {\bibfnamefont {N.~S.}\ \bibnamefont
  {Headings}}, \bibinfo {author} {\bibfnamefont {S.~M.}\ \bibnamefont
  {Hayden}}, \bibinfo {author} {\bibfnamefont {R.}~\bibnamefont {Coldea}}, \
  and\ \bibinfo {author} {\bibfnamefont {T.~G.}\ \bibnamefont {Perring}},\
  }\bibfield  {title} {\enquote {\bibinfo {title} {{Anomalous High-Energy Spin
  Excitations in the High-${T}_{c}$ Superconductor-Parent Antiferromagnet
  $\mathrm{La}_2\mathrm{CuO}_4$}},}\ }\href {\doibase
  10.1103/PhysRevLett.105.247001} {\bibfield  {journal} {\bibinfo  {journal}
  {Phys. Rev. Lett.}\ }\textbf {\bibinfo {volume} {105}},\ \bibinfo {pages}
  {247001} (\bibinfo {year} {2010})}\BibitemShut {NoStop}%
\bibitem [{\citenamefont {Martinelli}\ \emph {et~al.}(2022)\citenamefont
  {Martinelli}, \citenamefont {Betto}, \citenamefont {Kummer}, \citenamefont
  {Arpaia}, \citenamefont {Braicovich}, \citenamefont {Di~Castro},
  \citenamefont {Brookes}, \citenamefont {Moretti~Sala},\ and\ \citenamefont
  {Ghiringhelli}}]{Mar22}%
  \BibitemOpen
  \bibfield  {author} {\bibinfo {author} {\bibfnamefont {Leonardo}\
  \bibnamefont {Martinelli}}, \bibinfo {author} {\bibfnamefont {Davide}\
  \bibnamefont {Betto}}, \bibinfo {author} {\bibfnamefont {Kurt}\ \bibnamefont
  {Kummer}}, \bibinfo {author} {\bibfnamefont {Riccardo}\ \bibnamefont
  {Arpaia}}, \bibinfo {author} {\bibfnamefont {Lucio}\ \bibnamefont
  {Braicovich}}, \bibinfo {author} {\bibfnamefont {Daniele}\ \bibnamefont
  {Di~Castro}}, \bibinfo {author} {\bibfnamefont {Nicholas~B.}\ \bibnamefont
  {Brookes}}, \bibinfo {author} {\bibfnamefont {Marco}\ \bibnamefont
  {Moretti~Sala}}, \ and\ \bibinfo {author} {\bibfnamefont {Giacomo}\
  \bibnamefont {Ghiringhelli}},\ }\bibfield  {title} {\enquote {\bibinfo
  {title} {{Fractional Spin Excitations in the Infinite-Layer Cuprate
  $\mathrm{CaCuO}_2$}},}\ }\href {\doibase 10.1103/PhysRevX.12.021041}
  {\bibfield  {journal} {\bibinfo  {journal} {Phys. Rev. X}\ }\textbf {\bibinfo
  {volume} {12}},\ \bibinfo {pages} {021041} (\bibinfo {year}
  {2022})}\BibitemShut {NoStop}%
\bibitem [{\citenamefont {Tsyrulin}\ \emph {et~al.}(2009)\citenamefont
  {Tsyrulin}, \citenamefont {Pardini}, \citenamefont {Singh}, \citenamefont
  {Xiao}, \citenamefont {Link}, \citenamefont {Schneidewind}, \citenamefont
  {Hiess}, \citenamefont {Landee}, \citenamefont {Turnbull},\ and\
  \citenamefont {Kenzelmann}}]{Tsy09}%
  \BibitemOpen
  \bibfield  {author} {\bibinfo {author} {\bibfnamefont {N.}~\bibnamefont
  {Tsyrulin}}, \bibinfo {author} {\bibfnamefont {T.}~\bibnamefont {Pardini}},
  \bibinfo {author} {\bibfnamefont {R.~R.~P.}\ \bibnamefont {Singh}}, \bibinfo
  {author} {\bibfnamefont {F.}~\bibnamefont {Xiao}}, \bibinfo {author}
  {\bibfnamefont {P.}~\bibnamefont {Link}}, \bibinfo {author} {\bibfnamefont
  {A.}~\bibnamefont {Schneidewind}}, \bibinfo {author} {\bibfnamefont
  {A.}~\bibnamefont {Hiess}}, \bibinfo {author} {\bibfnamefont {C.~P.}\
  \bibnamefont {Landee}}, \bibinfo {author} {\bibfnamefont {M.~M.}\
  \bibnamefont {Turnbull}}, \ and\ \bibinfo {author} {\bibfnamefont
  {M.}~\bibnamefont {Kenzelmann}},\ }\bibfield  {title} {\enquote {\bibinfo
  {title} {{Quantum Effects in a Weakly Frustrated $\mathit{S}=1/2$
  Two-Dimensional Heisenberg Antiferromagnet in an Applied Magnetic Field}},}\
  }\href {\doibase 10.1103/PhysRevLett.102.197201} {\bibfield  {journal}
  {\bibinfo  {journal} {Phys. Rev. Lett.}\ }\textbf {\bibinfo {volume} {102}},\
  \bibinfo {pages} {197201} (\bibinfo {year} {2009})}\BibitemShut {NoStop}%
\bibitem [{\citenamefont {Tsyrulin}\ \emph {et~al.}(2010)\citenamefont
  {Tsyrulin}, \citenamefont {Xiao}, \citenamefont {Schneidewind}, \citenamefont
  {Link}, \citenamefont {R\o{}nnow}, \citenamefont {Gavilano}, \citenamefont
  {Landee}, \citenamefont {Turnbull},\ and\ \citenamefont
  {Kenzelmann}}]{Tsy10}%
  \BibitemOpen
  \bibfield  {author} {\bibinfo {author} {\bibfnamefont {N.}~\bibnamefont
  {Tsyrulin}}, \bibinfo {author} {\bibfnamefont {F.}~\bibnamefont {Xiao}},
  \bibinfo {author} {\bibfnamefont {A.}~\bibnamefont {Schneidewind}}, \bibinfo
  {author} {\bibfnamefont {P.}~\bibnamefont {Link}}, \bibinfo {author}
  {\bibfnamefont {H.~M.}\ \bibnamefont {R\o{}nnow}}, \bibinfo {author}
  {\bibfnamefont {J.}~\bibnamefont {Gavilano}}, \bibinfo {author}
  {\bibfnamefont {C.~P.}\ \bibnamefont {Landee}}, \bibinfo {author}
  {\bibfnamefont {M.~M.}\ \bibnamefont {Turnbull}}, \ and\ \bibinfo {author}
  {\bibfnamefont {M.}~\bibnamefont {Kenzelmann}},\ }\bibfield  {title}
  {\enquote {\bibinfo {title} {{Two-Dimensional Square-Lattice
  $\mathit{S}=\frac{1}{2}$ Antiferromagnet
  $\mathrm{Cu}{(\mathrm{pz})}_{2}{({\mathrm{ClO}}_{4})}_{2}$}},}\ }\href
  {\doibase 10.1103/PhysRevB.81.134409} {\bibfield  {journal} {\bibinfo
  {journal} {Phys. Rev. B}\ }\textbf {\bibinfo {volume} {81}},\ \bibinfo
  {pages} {134409} (\bibinfo {year} {2010})}\BibitemShut {NoStop}%
\bibitem [{\citenamefont {Powalski}\ \emph {et~al.}(2018)\citenamefont
  {Powalski}, \citenamefont {Schmidt},\ and\ \citenamefont {Uhrig}}]{Pow18}%
  \BibitemOpen
  \bibfield  {author} {\bibinfo {author} {\bibfnamefont {Michael}\ \bibnamefont
  {Powalski}}, \bibinfo {author} {\bibfnamefont {Kai~P.}\ \bibnamefont
  {Schmidt}}, \ and\ \bibinfo {author} {\bibfnamefont {Götz~S.}\ \bibnamefont
  {Uhrig}},\ }\bibfield  {title} {\enquote {\bibinfo {title} {Mutually
  attracting spin waves in the square-lattice quantum antiferromagnet},}\
  }\href {\doibase 10.21468/SciPostPhys.4.1.001} {\bibfield  {journal}
  {\bibinfo  {journal} {SciPost Phys.}\ }\textbf {\bibinfo {volume} {4}},\
  \bibinfo {pages} {001} (\bibinfo {year} {2018})}\BibitemShut {NoStop}%
\bibitem [{\citenamefont {Chen}\ and\ \citenamefont {Levy}(1971)}]{Che71}%
  \BibitemOpen
  \bibfield  {author} {\bibinfo {author} {\bibfnamefont {H.~H.}\ \bibnamefont
  {Chen}}\ and\ \bibinfo {author} {\bibfnamefont {Peter~M.}\ \bibnamefont
  {Levy}},\ }\bibfield  {title} {\enquote {\bibinfo {title} {{Quadrupole Phase
  Transitions in Magnetic Solids}},}\ }\href {\doibase
  10.1103/PhysRevLett.27.1383} {\bibfield  {journal} {\bibinfo  {journal}
  {Phys. Rev. Lett.}\ }\textbf {\bibinfo {volume} {27}},\ \bibinfo {pages}
  {1383–1385} (\bibinfo {year} {1971})}\BibitemShut {NoStop}%
\bibitem [{\citenamefont {Chandra}\ and\ \citenamefont
  {Coleman}(1991)}]{Cha91}%
  \BibitemOpen
  \bibfield  {author} {\bibinfo {author} {\bibfnamefont {P.}~\bibnamefont
  {Chandra}}\ and\ \bibinfo {author} {\bibfnamefont {P.}~\bibnamefont
  {Coleman}},\ }\bibfield  {title} {\enquote {\bibinfo {title} {{Quantum spin
  nematics: Moment-free magnetism}},}\ }\href {\doibase
  10.1103/PhysRevLett.66.100} {\bibfield  {journal} {\bibinfo  {journal} {Phys.
  Rev. Lett.}\ }\textbf {\bibinfo {volume} {66}},\ \bibinfo {pages} {100–103}
  (\bibinfo {year} {1991})}\BibitemShut {NoStop}%
\bibitem [{\citenamefont {L\"auchli}\ \emph {et~al.}(2006)\citenamefont
  {L\"auchli}, \citenamefont {Mila},\ and\ \citenamefont {Penc}}]{Lau06}%
  \BibitemOpen
  \bibfield  {author} {\bibinfo {author} {\bibfnamefont {Andreas}\ \bibnamefont
  {L\"auchli}}, \bibinfo {author} {\bibfnamefont {Fr\'ed\'eric}\ \bibnamefont
  {Mila}}, \ and\ \bibinfo {author} {\bibfnamefont {Karlo}\ \bibnamefont
  {Penc}},\ }\bibfield  {title} {\enquote {\bibinfo {title} {{Quadrupolar
  Phases of the $S=1$ Bilinear-Biquadratic Heisenberg Model on the Triangular
  Lattice}},}\ }\href {\doibase 10.1103/PhysRevLett.97.087205} {\bibfield
  {journal} {\bibinfo  {journal} {Phys. Rev. Lett.}\ }\textbf {\bibinfo
  {volume} {97}},\ \bibinfo {pages} {087205} (\bibinfo {year}
  {2006})}\BibitemShut {NoStop}%
\bibitem [{\citenamefont {Zhitomirsky}\ and\ \citenamefont
  {Tsunetsugu}(2010)}]{Zhi10}%
  \BibitemOpen
  \bibfield  {author} {\bibinfo {author} {\bibfnamefont {M.~E.}\ \bibnamefont
  {Zhitomirsky}}\ and\ \bibinfo {author} {\bibfnamefont {H.}~\bibnamefont
  {Tsunetsugu}},\ }\bibfield  {title} {\enquote {\bibinfo {title} {{Magnon
  pairing in quantum spin nematic}},}\ }\href {\doibase
  10.1209/0295-5075/92/37001} {\bibfield  {journal} {\bibinfo  {journal} {EPL}\
  }\textbf {\bibinfo {volume} {92}},\ \bibinfo {pages} {37001} (\bibinfo {year}
  {2010})}\BibitemShut {NoStop}%
\bibitem [{\citenamefont {Witczak-Krempa}\ \emph {et~al.}(2014)\citenamefont
  {Witczak-Krempa}, \citenamefont {Chen}, \citenamefont {Kim},\ and\
  \citenamefont {Balents}}]{Wit14}%
  \BibitemOpen
  \bibfield  {author} {\bibinfo {author} {\bibfnamefont {William}\ \bibnamefont
  {Witczak-Krempa}}, \bibinfo {author} {\bibfnamefont {Gang}\ \bibnamefont
  {Chen}}, \bibinfo {author} {\bibfnamefont {Yong~Baek}\ \bibnamefont {Kim}}, \
  and\ \bibinfo {author} {\bibfnamefont {Leon}\ \bibnamefont {Balents}},\
  }\bibfield  {title} {\enquote {\bibinfo {title} {{Correlated Quantum
  Phenomena in the Strong Spin-Orbit Regime}},}\ }\href {\doibase
  10.1146/annurev-conmatphys-020911-125138} {\bibfield  {journal} {\bibinfo
  {journal} {Annu. Rev. Condens. Matter Phys.}\ }\textbf {\bibinfo {volume}
  {5}},\ \bibinfo {pages} {57--82} (\bibinfo {year} {2014})}\BibitemShut
  {NoStop}%
\bibitem [{\citenamefont {Kim}\ \emph {et~al.}(2012)\citenamefont {Kim},
  \citenamefont {Casa}, \citenamefont {Upton}, \citenamefont {Gog},
  \citenamefont {Kim}, \citenamefont {Mitchell}, \citenamefont
  {Van~Veenendaal}, \citenamefont {Daghofer}, \citenamefont {Van Den~Brink},
  \citenamefont {Khaliullin},\ and\ \citenamefont {Kim}}]{Kim12}%
  \BibitemOpen
  \bibfield  {author} {\bibinfo {author} {\bibfnamefont {Jungho}\ \bibnamefont
  {Kim}}, \bibinfo {author} {\bibfnamefont {D.}~\bibnamefont {Casa}}, \bibinfo
  {author} {\bibfnamefont {M.~H.}\ \bibnamefont {Upton}}, \bibinfo {author}
  {\bibfnamefont {T.}~\bibnamefont {Gog}}, \bibinfo {author} {\bibfnamefont
  {Young-June}\ \bibnamefont {Kim}}, \bibinfo {author} {\bibfnamefont {J.~F.}\
  \bibnamefont {Mitchell}}, \bibinfo {author} {\bibfnamefont {M.}~\bibnamefont
  {Van~Veenendaal}}, \bibinfo {author} {\bibfnamefont {M.}~\bibnamefont
  {Daghofer}}, \bibinfo {author} {\bibfnamefont {J.}~\bibnamefont {Van
  Den~Brink}}, \bibinfo {author} {\bibfnamefont {G.}~\bibnamefont
  {Khaliullin}}, \ and\ \bibinfo {author} {\bibfnamefont {B.~J.}\ \bibnamefont
  {Kim}},\ }\bibfield  {title} {\enquote {\bibinfo {title} {{Magnetic
  Excitation Spectra of $\mathrm{Sr_2IrO_4}$ Probed by Resonant Inelastic X-Ray
  Scattering: Establishing Links to Cuprate Superconductors}},}\ }\href
  {\doibase 10.1103/PhysRevLett.108.177003} {\bibfield  {journal} {\bibinfo
  {journal} {Phys. Rev. Lett.}\ }\textbf {\bibinfo {volume} {108}},\ \bibinfo
  {pages} {177003} (\bibinfo {year} {2012})}\BibitemShut {NoStop}%
\bibitem [{\citenamefont {McMorrow}\ \emph {et~al.}(2001)\citenamefont
  {McMorrow}, \citenamefont {McEwen}, \citenamefont {Steigenberger},
  \citenamefont {R\o{}nnow},\ and\ \citenamefont {Yakhou}}]{Mcm01}%
  \BibitemOpen
  \bibfield  {author} {\bibinfo {author} {\bibfnamefont {D.~F.}\ \bibnamefont
  {McMorrow}}, \bibinfo {author} {\bibfnamefont {K.~A.}\ \bibnamefont
  {McEwen}}, \bibinfo {author} {\bibfnamefont {U.}~\bibnamefont
  {Steigenberger}}, \bibinfo {author} {\bibfnamefont {H.~M.}\ \bibnamefont
  {R\o{}nnow}}, \ and\ \bibinfo {author} {\bibfnamefont {F.}~\bibnamefont
  {Yakhou}},\ }\bibfield  {title} {\enquote {\bibinfo {title} {{X-Ray Resonant
  Scattering Study of the Quadrupolar Order in ${\mathrm{UPd}}_{3}$}},}\ }\href
  {\doibase 10.1103/PhysRevLett.87.057201} {\bibfield  {journal} {\bibinfo
  {journal} {Phys. Rev. Lett.}\ }\textbf {\bibinfo {volume} {87}},\ \bibinfo
  {pages} {057201} (\bibinfo {year} {2001})}\BibitemShut {NoStop}%
\bibitem [{\citenamefont {Mourigal}\ \emph {et~al.}(2012)\citenamefont
  {Mourigal}, \citenamefont {Enderle}, \citenamefont {F\aa{}k}, \citenamefont
  {Kremer}, \citenamefont {Law}, \citenamefont {Schneidewind}, \citenamefont
  {Hiess},\ and\ \citenamefont {Prokofiev}}]{Mou12}%
  \BibitemOpen
  \bibfield  {author} {\bibinfo {author} {\bibfnamefont {M.}~\bibnamefont
  {Mourigal}}, \bibinfo {author} {\bibfnamefont {M.}~\bibnamefont {Enderle}},
  \bibinfo {author} {\bibfnamefont {B.}~\bibnamefont {F\aa{}k}}, \bibinfo
  {author} {\bibfnamefont {R.~K.}\ \bibnamefont {Kremer}}, \bibinfo {author}
  {\bibfnamefont {J.~M.}\ \bibnamefont {Law}}, \bibinfo {author} {\bibfnamefont
  {A.}~\bibnamefont {Schneidewind}}, \bibinfo {author} {\bibfnamefont
  {A.}~\bibnamefont {Hiess}}, \ and\ \bibinfo {author} {\bibfnamefont
  {A.}~\bibnamefont {Prokofiev}},\ }\bibfield  {title} {\enquote {\bibinfo
  {title} {{Evidence of a Bond-Nematic Phase in ${\mathrm{LiCuVO}}_{4}$}},}\
  }\href {\doibase 10.1103/PhysRevLett.109.027203} {\bibfield  {journal}
  {\bibinfo  {journal} {Phys. Rev. Lett.}\ }\textbf {\bibinfo {volume} {109}},\
  \bibinfo {pages} {027203} (\bibinfo {year} {2012})}\BibitemShut {NoStop}%
\bibitem [{\citenamefont {Seyler}\ \emph {et~al.}(2020)\citenamefont {Seyler},
  \citenamefont {de~la Torre}, \citenamefont {Porter}, \citenamefont {Zoghlin},
  \citenamefont {Polski}, \citenamefont {Nguyen}, \citenamefont {Nadj-Perge},
  \citenamefont {Wilson},\ and\ \citenamefont {Hsieh}}]{Sey20}%
  \BibitemOpen
  \bibfield  {author} {\bibinfo {author} {\bibfnamefont {K.~L.}\ \bibnamefont
  {Seyler}}, \bibinfo {author} {\bibfnamefont {A.}~\bibnamefont {de~la Torre}},
  \bibinfo {author} {\bibfnamefont {Z.}~\bibnamefont {Porter}}, \bibinfo
  {author} {\bibfnamefont {E.}~\bibnamefont {Zoghlin}}, \bibinfo {author}
  {\bibfnamefont {R.}~\bibnamefont {Polski}}, \bibinfo {author} {\bibfnamefont
  {M.}~\bibnamefont {Nguyen}}, \bibinfo {author} {\bibfnamefont
  {S.}~\bibnamefont {Nadj-Perge}}, \bibinfo {author} {\bibfnamefont {S.~D.}\
  \bibnamefont {Wilson}}, \ and\ \bibinfo {author} {\bibfnamefont
  {D.}~\bibnamefont {Hsieh}},\ }\bibfield  {title} {\enquote {\bibinfo {title}
  {{Spin-orbit-enhanced magnetic surface second-harmonic generation in
  ${\mathrm{Sr}}_{2}\mathrm{Ir}{\mathrm{O}}_{4}$}},}\ }\href {\doibase
  10.1103/PhysRevB.102.201113} {\bibfield  {journal} {\bibinfo  {journal}
  {Phys. Rev. B}\ }\textbf {\bibinfo {volume} {102}},\ \bibinfo {pages}
  {201113} (\bibinfo {year} {2020})}\BibitemShut {NoStop}%
\bibitem [{\citenamefont {Jeong}\ \emph {et~al.}(2017)\citenamefont {Jeong},
  \citenamefont {Sidis}, \citenamefont {Louat}, \citenamefont {Brouet},\ and\
  \citenamefont {Bourges}}]{Jeo17}%
  \BibitemOpen
  \bibfield  {author} {\bibinfo {author} {\bibfnamefont {Jaehong}\ \bibnamefont
  {Jeong}}, \bibinfo {author} {\bibfnamefont {Yvan}\ \bibnamefont {Sidis}},
  \bibinfo {author} {\bibfnamefont {Alex}\ \bibnamefont {Louat}}, \bibinfo
  {author} {\bibfnamefont {Véronique}\ \bibnamefont {Brouet}}, \ and\ \bibinfo
  {author} {\bibfnamefont {Philippe}\ \bibnamefont {Bourges}},\ }\bibfield
  {title} {\enquote {\bibinfo {title} {{Time-reversal Symmetry Breaking Hidden
  Order in Sr2(Ir,Rh)O4}},}\ }\href {\doibase 10.1038/ncomms15119} {\bibfield
  {journal} {\bibinfo  {journal} {Nat. Commun.}\ }\textbf {\bibinfo {volume}
  {8}},\ \bibinfo {pages} {15119} (\bibinfo {year} {2017})}\BibitemShut
  {NoStop}%
\bibitem [{\citenamefont {Murayama}\ \emph {et~al.}(2021)\citenamefont
  {Murayama}, \citenamefont {Ishida}, \citenamefont {Kurihara}, \citenamefont
  {Ono}, \citenamefont {Sato}, \citenamefont {Kasahara}, \citenamefont
  {Watanabe}, \citenamefont {Yanase}, \citenamefont {Cao}, \citenamefont
  {Mizukami}, \citenamefont {Shibauchi}, \citenamefont {Matsuda},\ and\
  \citenamefont {Kasahara}}]{Mur21}%
  \BibitemOpen
  \bibfield  {author} {\bibinfo {author} {\bibfnamefont {H.}~\bibnamefont
  {Murayama}}, \bibinfo {author} {\bibfnamefont {K.}~\bibnamefont {Ishida}},
  \bibinfo {author} {\bibfnamefont {R.}~\bibnamefont {Kurihara}}, \bibinfo
  {author} {\bibfnamefont {T.}~\bibnamefont {Ono}}, \bibinfo {author}
  {\bibfnamefont {Y.}~\bibnamefont {Sato}}, \bibinfo {author} {\bibfnamefont
  {Y.}~\bibnamefont {Kasahara}}, \bibinfo {author} {\bibfnamefont
  {H.}~\bibnamefont {Watanabe}}, \bibinfo {author} {\bibfnamefont
  {Y.}~\bibnamefont {Yanase}}, \bibinfo {author} {\bibfnamefont
  {G.}~\bibnamefont {Cao}}, \bibinfo {author} {\bibfnamefont {Y.}~\bibnamefont
  {Mizukami}}, \bibinfo {author} {\bibfnamefont {T.}~\bibnamefont {Shibauchi}},
  \bibinfo {author} {\bibfnamefont {Y.}~\bibnamefont {Matsuda}}, \ and\
  \bibinfo {author} {\bibfnamefont {S.}~\bibnamefont {Kasahara}},\ }\bibfield
  {title} {\enquote {\bibinfo {title} {{Bond Directional Anapole Order in a
  Spin-Orbit Coupled Mott Insulator
  ${\mathrm{Sr}}_{2}({\mathrm{Ir}}_{1\ensuremath{-}x}{\mathrm{Rh}}_{x}){\mathrm{O}}_{4}$}},}\
  }\href {\doibase 10.1103/PhysRevX.11.011021} {\bibfield  {journal} {\bibinfo
  {journal} {Phys. Rev. X}\ }\textbf {\bibinfo {volume} {11}},\ \bibinfo
  {pages} {011021} (\bibinfo {year} {2021})}\BibitemShut {NoStop}%
\bibitem [{\citenamefont {Kim}\ \emph {et~al.}(2022)\citenamefont {Kim},
  \citenamefont {Kim}, \citenamefont {Kim}, \citenamefont {Park}, \citenamefont
  {Kim}, \citenamefont {Kwon}, \citenamefont {Kim}, \citenamefont {Seo},
  \citenamefont {Kim},\ and\ \citenamefont {Kim}}]{kim22}%
  \BibitemOpen
  \bibfield  {author} {\bibinfo {author} {\bibfnamefont {Jimin}\ \bibnamefont
  {Kim}}, \bibinfo {author} {\bibfnamefont {Hoon}\ \bibnamefont {Kim}},
  \bibinfo {author} {\bibfnamefont {Hyun-Woo~J.}\ \bibnamefont {Kim}}, \bibinfo
  {author} {\bibfnamefont {Sunwook}\ \bibnamefont {Park}}, \bibinfo {author}
  {\bibfnamefont {Jin-Kwang}\ \bibnamefont {Kim}}, \bibinfo {author}
  {\bibfnamefont {Junyoung}\ \bibnamefont {Kwon}}, \bibinfo {author}
  {\bibfnamefont {Jungho}\ \bibnamefont {Kim}}, \bibinfo {author}
  {\bibfnamefont {Hyeong~Woo}\ \bibnamefont {Seo}}, \bibinfo {author}
  {\bibfnamefont {Jun~Sung}\ \bibnamefont {Kim}}, \ and\ \bibinfo {author}
  {\bibfnamefont {B.~J.}\ \bibnamefont {Kim}},\ }\bibfield  {title} {\enquote
  {\bibinfo {title} {Single crystal growth of iridates without platinum
  impurities},}\ }\href {\doibase 10.1103/PhysRevMaterials.6.103401} {\bibfield
   {journal} {\bibinfo  {journal} {Phys. Rev. Mater.}\ }\textbf {\bibinfo
  {volume} {6}},\ \bibinfo {pages} {103401} (\bibinfo {year}
  {2022})}\BibitemShut {NoStop}%
\bibitem [{\citenamefont {Torchinsky}\ \emph {et~al.}(2015)\citenamefont
  {Torchinsky}, \citenamefont {Chu}, \citenamefont {Zhao}, \citenamefont
  {Perkins}, \citenamefont {Sizyuk}, \citenamefont {Qi}, \citenamefont {Cao},\
  and\ \citenamefont {Hsieh}}]{Tor15_SHG}%
  \BibitemOpen
  \bibfield  {author} {\bibinfo {author} {\bibfnamefont {D.}~\bibnamefont
  {Torchinsky}}, \bibinfo {author} {\bibfnamefont {H.}~\bibnamefont {Chu}},
  \bibinfo {author} {\bibfnamefont {L.}~\bibnamefont {Zhao}}, \bibinfo {author}
  {\bibfnamefont {N.~B}\ \bibnamefont {Perkins}}, \bibinfo {author}
  {\bibfnamefont {Y.}~\bibnamefont {Sizyuk}}, \bibinfo {author} {\bibfnamefont
  {T.}~\bibnamefont {Qi}}, \bibinfo {author} {\bibfnamefont {G.}~\bibnamefont
  {Cao}}, \ and\ \bibinfo {author} {\bibfnamefont {D.}~\bibnamefont {Hsieh}},\
  }\bibfield  {title} {\enquote {\bibinfo {title} {{Structural Distortion-Induced Magnetoelastic Locking in ${\mathrm{Sr}}_{2}{\mathrm{IrO}}_{4}$ Revealed through Nonlinear Optical
  Harmonic Generation}},}\ }\href {\doibase 10.1103/PhysRevLett.114.096404}
  {\bibfield  {journal} {\bibinfo  {journal} {Phys. Rev. Lett.}\ }\textbf
  {\bibinfo {volume} {114}},\ \bibinfo {pages} {096404} (\bibinfo {year}
  {2015})}\BibitemShut {NoStop}%
\bibitem [{\citenamefont {Zhu}\ \emph {et~al.}(2021)\citenamefont {Zhu},
  \citenamefont {Ullah},\ and\ \citenamefont {Taufour}}]{Zhu21}%
  \BibitemOpen
  \bibfield  {author} {\bibinfo {author} {\bibfnamefont {X.~D.}\ \bibnamefont
  {Zhu}}, \bibinfo {author} {\bibfnamefont {R.}~\bibnamefont {Ullah}}, \ and\
  \bibinfo {author} {\bibfnamefont {V.}~\bibnamefont {Taufour}},\ }\bibfield
  {title} {\enquote {\bibinfo {title} {{Oblique-incidence Sagnac
  interferometric scanning microscope for studying magneto-optic effects of
  materials at low temperatures}},}\ }\href {\doibase 10.1063/5.0042574}
  {\bibfield  {journal} {\bibinfo  {journal} {Rev. Sci. Instrum.}\ }\textbf
  {\bibinfo {volume} {92}},\ \bibinfo {pages} {043706} (\bibinfo {year}
  {2021})}\BibitemShut {NoStop}%
\end{thebibliography}
%

\section*{Methods}
\textbf{Crystal Growth} Single crystals of Sr$_2$IrO$_4$ were grown by the standard flux growth method. Powders of IrO$_2$, SrCO$_3$, and SrCl $_2 \cdot $ 6H$_2$O were mixed and placed in an iridium crucible covered with a lid. The mixture was melted and soaked at $T$ = 1300$^\circ$C, slowly cooled down to 900$^\circ$C at 8$^\circ$C$/$h, and then furnace cooled to room temperature. We note that our crystals grown in an iridium crucible have the lattice structure of space group $I4_1/acd$ (ref.~\citenum{kim22}) and shows no sign of glide symmetry breaking distortions previously reported~\cite{Tor15_SHG}.

\noindent
\textbf{Resonant x-ray diffraction (RXD)} RXD experiments were carried out at the 1C beamline of Pohang Accelerator Laboratory and 4-ID-D beamline of the Advanced Photon Source. Incident x-ray was tuned at the Ir $L_3$ edge (11.217 keV). The focused beam having spatial resolution better than $\sim$ 100 $\mu$m was used. The sample was mounted on the cold finger of a closed-cycle cryostat and temperature was kept at 10 K throughout the experiment. The left and right circular polarized x-ray was generated using a diamond phase retarder. The helicity was switched at every data point to measure the flipping ratio of circular dichroic signal.

\noindent
\textbf{Resonant inelastic x-ray sacttering} RIXS spectra were measured at the 27-ID-B beamline of the Advanced Photon Source. Incident x-ray was tuned to the Ir $L_3$ edge (11.215 keV). Using a diamond~(1\,1\,1) high-heat-load monochromator in combination with a Si~(8\,4\,4) channel-cut monochromator reduced the energy bandpass down to 14.8\,meV. The beam was then focused by a set of Kirkpatrick-Baez mirrors, producing a spot size of $40\times10$ (H$\times$V)\,\textmu m$^2$ FWHM at the sample position. Scattered photons were analyzed by a Si~(8\,4\,4) diced spherical analyzer with a radius of 2\,m and with a mask of 2\,cm diameter. The overall energy resolution was about 30\,meV. A horizontal scattering geometry was used with the incident $\pi$-polarization and the outgoing polarization was not resolved. All RIXS spectra were taken around (3\,0\,28.5) in a normal incident scattering geometry and a small magnetic field ($\sim0.3$\,T) was applied along the either $a$- or $b$-axis to align magnetic moments. The scattering angle (2$\theta$) was kept around 90$^{\circ}$ to suppress elastic Thomson scatterings.

\noindent
\textbf{Raman spectroscopy} Raman spectroscopy was performed with a home-built setup equipped with a 633\,nm He-Ne laser and a liquid nitrogen cooled CCD (Princeton instruments). The elastic signal is removed by grating-based notch filters (Optigrate, BragGrate$^{\textrm{TM}}$ Notch filters). The spectra were acquired on as-grown surfaces of Sr$_2$IrO$_4$ crystals mounted in a closed cycle optical cryostat (Montana instruments). The laser power and beam spot size are 0.8\,mW and 2\,$\mu$m, respectively, which resulted in an almost temperature independent laser heating of 25\,K as determined from the Stokes to Anti-Stokes intensity ratio. All Raman spectra are Bose-corrected.

\noindent
\textbf{Magneto-optical Kerr measurement} We performed magneto-optical Kerr measurements on Sr$_2$IrO$_4$ using an oblique-incidence zero-area Sagnac interferometer operating at 1550\,nm wavelength to measure the in-plane magnetization~\cite{Zhu21}. The relative Kerr angle $\Delta\theta_K$ is obtained by subtracting temperature-independent backgrounds, coming from the instrumentation offset, measured at $T$\,=\,300\,K, and converted to magnetization (in $\mu_B$/ion) by a quantitative comparison of our 0.35\,T Kerr data with the corresponding magnetization data in the Ref.~\citenum{Por19}. Throughout the measurements, the incident optical power was maintained below 1\,mW so that the effect of optical heating was smaller than 1\,K.

\section*{Acknowledgments}
We thank Nic Shannon, Giniyat Khaliullin and Yong Baek Kim for helpful discussions. This project is supported by IBS-R014-A2 and and National Research Foundation (NRF) of Korea through the SRC (No.~2018R1A5A6075964). The use of the Advanced Photon Source at the Argonne National Laboratory was supported by the U.~S.~DOE under Contract No.~DE-AC02-06CH11357. G.Y.C. is supported by the NRF of Korea (Grant No.~2020R1C1C1006048) funded by the Korean Government (MSIT) as well as the Institute of Basic Science under project code IBS-R014-D1. G.Y.C. is also supported by the Air Force Office of Scientific Research under Award No.~FA2386-22-1-4061 and G.Y.C. also acknowledges Samsung Science and Technology Foundation under Project Number SSTF-BA2002-05. H.H. and J.J. are supported by the National Research Foundation of Korea (Grant No.~2020R1A5A1016518), and the Creative-Pioneering Researchers Program through Seoul National University. 

\section*{Authors contributions} 
B.J.K. conceived and managed the project. H.K., K.K. and Jo.K. performed Raman experiments. H.K., S.H., J.S., G.F., Y.C., D.H. and J.W.K. performed RXD experiments. H.K., J.-K.K., H.-W.J.K. and Ju.K. performed RIXS experiments and analyzed the data. Ji.K. grew single crystals. H.H. and J.J. performed Kerr measurements. H.K. and B.J.K. performed representation analysis. W.L. and G.Y.C. assisted interpretation of the data.  H.K., J.-K.K. and B.J.K. wrote the manuscript with inputs from all authors. 

\section*{Competing interests} 
The authors declare no competing interests. 

\section*{Data and materials availability}
All data is available in the manuscript or the supplementary information.

\setcounter{figure}{0}
\setcounter{table}{0}
\makeatletter
\renewcommand{\fnum@figure}{{\bf Extended Data Fig.\:\thefigure}}
\renewcommand{\fnum@table}{{\bf Extended Data Table \thetable}}
\@fpsep\textheight
\makeatother

\begin{figure*}[t!]
\centering
\includegraphics[width = 0.7\textwidth]{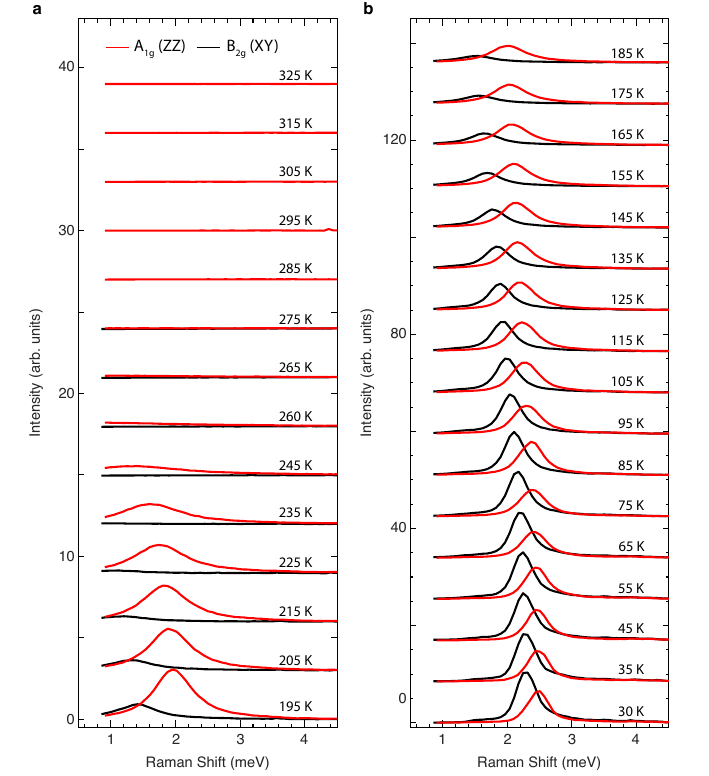}
\caption{\label{fig:ed1}{\bf Temperature evolution of the low-energy Raman modes.} The Raman spectra of the $A_{\rm 1g}$ and $B_{\rm 2g}$ modes shown in Figs.~\ref{fig:2}c and~\ref{fig:2}e are displayed with vertical offset for clarity.  The $A_{\rm 1g}$ (red) and $B_{\rm 2g}$ (black) spectra were measured on the same crystal under the same experimental conditions including laser power and acquisition time, and the spectra are plotted in the same arbitrary unit.}
\end{figure*}

\begin{figure*}[hbt!]
\centering
\includegraphics[width = 0.7\textwidth]{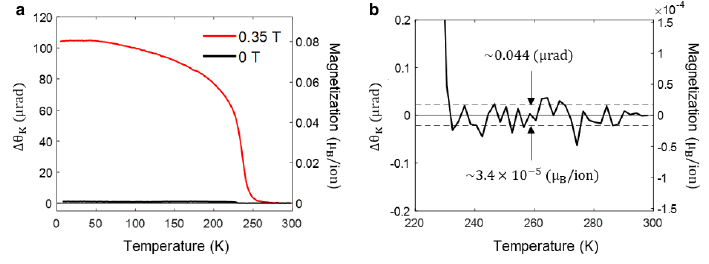}
\caption{\label{fig:ed2}{\bf Magneto-optical Kerr measurement.} {\bf a,} Relative Kerr angle in the 0.35\,T in-plane magnetic field (red line) and ambient magnetic field near 0\,T (black line). The relative Kerr angle (left axis) is converted to magnetization (right axis) using a conversion factor of 7.7$\times 10^{-4}$ ($\mu_B$/Ir ion)/(µrad). {\bf b,} Magnified plot of the 0 \,T data in \textbf{a}. Dashed line indicates the standard deviation for the $T$\,=\,230$\,\sim\,$300\,K range. The Kerr signal at $B$\,=\,0\,T in the range of 230\,K\,$<\,T\,<\,$300$\,$K shows that no net magnetization is present within our experimental resolution, ~\,3.4$\times 10^{-5}\,\mu_B$/ion (dashed line), thus confirming the preservation of time-reversal symmetry above $T_{\rm N}\,\sim\,$230\,K.}
\end{figure*}

\begin{table*}[hbt!]
\caption{\label{tab:ed1} Sensitive mode for each combination of specific scattering geometry and field direction in RIXS measurements on Sr$_2$IrO$_4$. Here, L stands for the longitudinal mode, T for the in-plane transverse mode, and T$'$ for the out-of-plane transverse mode. Please note that `Mode' represents the most sensitive mode in a specific scattering geometry and a field direction. \\}
\begin{ruledtabular}
\begin{tabular}{cccccc}
 Incident angle & Field direction & Moment direction & Polarization & Sensitive element & Mode \\ \hline
\multirow{4}{*}{Normal} & \multirow{2}{*}{[100]} & \multirow{2}{*}{[010]} &                                   $e_{\pi}$$\times$$e_{\pi'}$ & [010] & L \\
& & & $e_{\pi}$$\times$$e_{\sigma'}$ & [001] & T$'$ \\ \cline{2-6}
                        & \multirow{2}{*}{[010]} & \multirow{2}{*}{[100]} & $e_{\pi}$$\times$$e_{\pi'}$ & [010] & T \\
& & & $e_{\pi}$$\times$$e_{\sigma'}$ & [001] & T$'$ \\ \cline{1-6}
\multirow{4}{*}{Grazing} & \multirow{2}{*}{[100]} & \multirow{2}{*}{[010]} &                                  $e_{\pi}$$\times$$e_{\pi'}$ & [010] & L \\
& & & $e_{\pi}$$\times$$e_{\sigma'}$ & [100] & T \\ \cline{2-6}
                        & \multirow{2}{*}{[010]} & \multirow{2}{*}{[100]} & $e_{\pi}$$\times$$e_{\pi'}$ & [010] & T \\
& & & $e_{\pi}$$\times$$e_{\sigma'}$ & [100] & L \\
\end{tabular}
\end{ruledtabular}
\end{table*}

\end{document}